\documentclass[conference]{IEEEtran}
\IEEEoverridecommandlockouts
\usepackage{cite}
\usepackage{amsmath,amssymb,amsfonts}
\usepackage{amsthm}
\usepackage{mathrsfs}

\newtheorem{corollary}{Corollary}
\newtheorem{definition}{Definition}

\usepackage[linesnumbered, ruled]{algorithm2e}
\usepackage{graphicx}
\usepackage{textcomp}
\usepackage{xcolor}
\usepackage[hyphens]{url}
\usepackage{fancyhdr}
\usepackage{booktabs}
\usepackage{verbatim}
\usepackage{hyperref}
\usepackage{subfigure}
\hypersetup{hidelinks}
\usepackage{setspace}
\usepackage{pifont}
\usepackage{etoolbox}
\usepackage{caption}
\usepackage{bbding}
\usepackage[square,sort,comma,numbers]{natbib}
\usepackage[misc]{ifsym}
\usepackage{multirow}
\usepackage{array}
\usepackage[bb=boondox,cal=boondox,scr=boondox]{mathalpha}
\def\BibTeX{{\rm B\kern-.05em{\sc i\kern-.025em b}\kern-.08em
    T\kern-.1667em\lower.7ex\hbox{E}\kern-.125emX}}

\IEEEoverridecommandlockouts\IEEEpubid{\makebox[\columnwidth]{979-8-3195-0662-7/26/\$31.00 $\copyright$2026 IEEE \hfill}\hspace{\columnsep}\makebox[\columnwidth]{ }}

\begin{document}
\bstctlcite{BSTcontrol}

\title{Beyond Callbacks: A Developer-friendly Programming Paradigm for Network Simulation}


\author{\IEEEauthorblockN{Yuanyi Zhu\textsuperscript{*}\thanks{\textsuperscript{*}Both authors contributed equally to this research.}}
\IEEEauthorblockA{\textit{Fudan University}} \\
\vspace{-1.0cm}
\and
\IEEEauthorblockN{Zijian Li\textsuperscript{*}}
\IEEEauthorblockA{\textit{Fudan University}} \\
\vspace{-1.0cm}
\and
\IEEEauthorblockN{Xin Ai}
\IEEEauthorblockA{\textit{Fudan University}} \\
\vspace{-1.0cm}
\and
\IEEEauthorblockN{Zixuan Chen}
\IEEEauthorblockA{\textit{China Telecom}} \\
\vspace{-1.0cm}
\and
\IEEEauthorblockN{Sen Liu}
\IEEEauthorblockA{\textit{Fudan University}} \\
\vspace{-1.0cm}
\and
\IEEEauthorblockN{Yang Xu\textsuperscript{$\dagger$}\thanks{\textsuperscript{$\dagger$}Corresponding Author}}
\IEEEauthorblockA{\textit{Fudan University}} \\
\vspace{-1.0cm}
}

\maketitle
\thispagestyle{plain} 

\begin{abstract}
Network simulation plays a crucial role in both networking research and industry. Existing commonly-used Discrete Event Simulations (DES) are based on the callback mechanism for discrete events. However, as network functions grow in complexity, callbacks reveal a fundamental structural limitation: a callback can only represent a single point in a blocking operation and cannot capture the full operation as a coherent whole. This forces developers to manually decompose sequential network logic into scattered callback fragments, leading to stack ripping, callback hell, and heavy development and debugging burdens.

We propose CoDES (\textbf{Co}routine-based \textbf{DES}), a coroutine-based programming paradigm for network simulation that addresses this structural limitation. CoDES targets development productivity rather than execution performance. Realizing CoDES in existing DES simulators is non-trivial, as coroutine suspension must be coordinated with the event scheduler, introducing challenges in coroutine resource conflicts, operation-level computation requirements, compatibility with existing callback-based simulators, and overhead. We implement a CoDES-based framework in ns-3, address all four challenges, and release it as open source.

We evaluate CoDES by implementing Message Passing Interface (MPI), High Precision Congestion Control (HPCC), and Routing Information Protocol (RIP) in ns-3, spanning the
transport, congestion control, and routing layers.
Compared with callback-based counterparts, CoDES reduces lines of code by up to 51.4\% and nesting depth by up to 50\%, with no accuracy loss and negligible runtime and memory overhead.

\end{abstract}


\section{Introduction}

Network simulation plays a pivotal role in the rapid advancement of network design and optimization. It is an essential tool for verifying and evaluating network designs, or in scenarios with limited experimental
resources \cite{m3}. Recent work on network simulation includes Unison \cite{UNISON} and DONS \cite{DONS}, which accelerate execution of network simulation through parallel technologies, and M3 \cite{m3}, a flow-level network simulator that leverages machine learning to accelerate simulation.
However, prior work focuses exclusively on execution performance, leaving the development burden unaddressed.

Current network simulators \cite{2008-ns3, opnet, omnet++, glomosim, hayamizu2019ceforesim} are predominantly based on Discrete Event Simulations (DES) and widely adopt callbacks to trigger and handle events.
Callback-based designs enable modular development by decoupling inter-module interactions, conforming to the Open/Closed Principle \cite{open-closed-principle} and improving the overall efficiency of network simulation development.

\begin{figure*}[htbp] 
\vspace{-1.0cm}
\centering
\setlength{\subfigcapskip}{-6pt}
\subfigure[Procedure A successfully gets data when data can be fetched immediately, e.g., local data fetch.]{
\label{fig:callback_bug_1}
\includegraphics[width=0.8\linewidth]{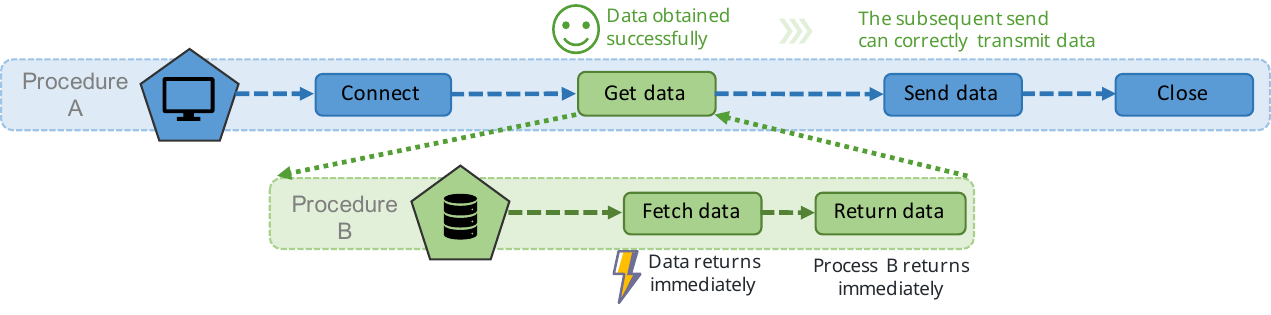}
}
\quad
\vspace{-0.2cm}

\subfigure[Procedure A fails to get data when data fetch involves blocking operations, e.g., remote data fetch.]{
\label{fig:callback_bug_2}
\includegraphics[width=0.8\linewidth]{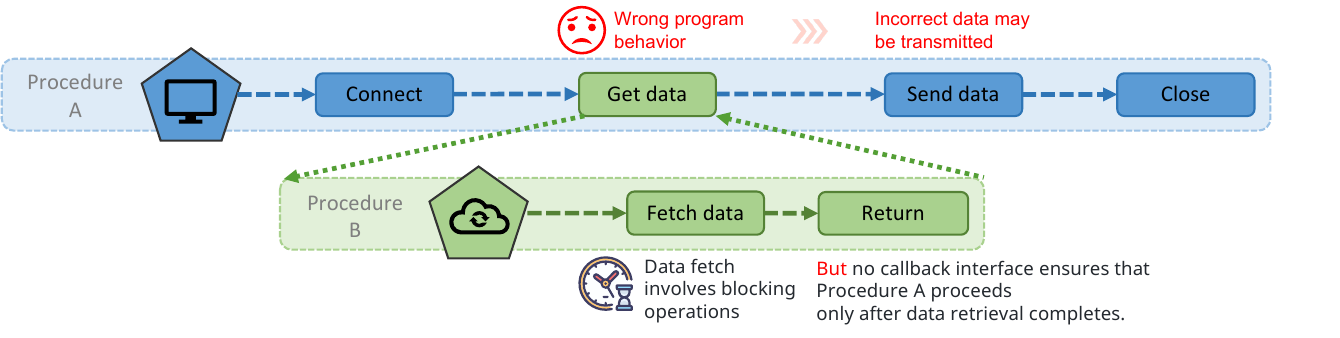}
}
\vspace{-0.3cm}
\caption{Data fetch and send example with network function extension.}
\label{fig:callback_bug}
\vspace{-0.5cm}
\end{figure*}

However, as network design and optimization functions grow more complex, the pervasive use of callbacks in network simulators reveals significant limitations at two levels: structural violations of design principles, and day-to-day development burdens.
At the structural level, callbacks depart from their original intent and violate the Open/Closed Principle. Consider two network components where Procedure A invokes Procedure B (Figure \ref{fig:callback_bug} and Section \ref{subsection:callback_problem}).
When a blocking operation is later introduced into B, B must expose an additional callback to notify A upon completion, which in turn forces A to update its own interface, a cascading effect that grows worse with system complexity.
This violates the Open/Closed Principle, since extending behavior requires modifying existing interfaces.
At the development level, callbacks impose substantial burdens on both development and debugging. Inverted control flow often forces developers to implement later callbacks first to satisfy earlier interfaces. Explicit management of callback state further distracts from core logic. As systems scale, callback chains become deeply nested, with more than 10 levels within a simulated node being common, leading to problems such as stack ripping \cite{stack-ripping} and callback hell
\cite{edwards2009callbackhell}.
These issues are not incidental: they reflect a deeper structural mismatch between how callbacks model network events and how those events actually behave.

\textbf{The fundamental reason is that callbacks cannot naturally simulate
blocking network events.}
In real-world programs, the OS transparently suspends and resumes blocking operations. In DES, however, a single callback can represent only one point in a blocking operation, its start or completion, but cannot capture the full operation as a coherent whole.
As a result, callbacks provide only a fragmented view of blocking behavior, creating mismatches between the sequential logic of real network programs and their simulations, and increasing the burden for programmers familiar with conventional network programming.

To reduce development effort and facilitate progress in network design and algorithms, a developer-friendly network simulator should follow the sequential control flow of real programs and provide a blocking-style interface for simulating blocking network operations.
This requires a simulation framework that supports lightweight suspension and resumption of network operations with automatic state management, incurring negligible overhead to accommodate the scale constraints of DES.
Fundamentally, this calls for a new programming paradigm for network simulation that expresses network operations as sequential code, rather than decomposed callbacks.

We propose CoDES (\textbf{Co}routine-based \textbf{DES}), a programming paradigm for network simulation built on C++20 stackless coroutines \cite{C++coroutine-2}.
The core insight is that a coroutine can suspend mid-execution at a blocking point and resume exactly where it left off, naturally expressing a blocking network operation as a single coherent function rather than splitting it across multiple callbacks.
Realizing this in a DES simulator is non-trivial, however, as coroutine suspension must be coordinated with the event scheduler, given that execution can only advance when the scheduler dispatches the next event.
CoDES integrates C++20 stackless coroutines into the existing simulators and provides bidirectional interoperability between callback and coroutine code, targeting development productivity rather than execution performance.
We implement a CoDES-based framework in ns-3, resolving the resulting system challenges in correctness, functionality, compatibility, and overhead, and release it as open source\footnote{\textbf{\href{https://github.com/CoDES-Development/CoDES}{https://github.com/CoDES-Development/CoDES}}}.
CoDES reduces code volume by up to 51.4\% and nesting depth by up to 50\% with negligible runtime overhead.


In summary, we outline three main contributions:
\begin{itemize}
  \item We point out that one key cause of the heavy development workloads in DES is
  callbacks' inability to naturally simulate blocking network events
  (Section \ref{sec:motivation}). We further demonstrate, both theoretically (Section \ref{sec:theory}) and empirically (Section \ref{sec:compatibility}), that callbacks and
  coroutines are interconvertible, yet differ markedly in how conveniently they
  express blocking network events. 

  \item We propose CoDES (\textbf{Co}routine-based
  \textbf{DES}), a programming paradigm for network simulation
  (Section \ref{sec:overview}), and implement a corresponding simulation framework
  leveraging the C++ coroutine mechanism \cite{C++coroutine-2}. 
  We further address key system
  challenges: coroutine resource conflicts, operation-level computation requirements,
  compatibility with existing callback-based simulators, and overhead 
  (Section \ref{sec:design}). We open-source our framework to promote adoption.

  \item We implement three representative network functions across layers, namely MPI
  \cite{walker1996mpi}, HPCC \cite{2019-sigcomm-HPCC}, and RIP \cite{stark1999rip}, in
  ns-3 based on CoDES (Section \ref{sec:implementation}). Compared with
  callback-based implementations, CoDES reduces LOC by up to 51.4\% and
  function-call nesting depth by up to 50\%, with no accuracy loss and
  negligible runtime and memory overhead (Section \ref{sec:evaluation}).
\end{itemize}

\section{Background \& Motivation}
\label{sec:motivation}






\begin{figure*}[htbp] 
\vspace{-1.0cm}
\centering

\subfigure[Real-world program of local data fetch and update to remote data fetch]{
\label{fig:callback_bug_code_realworld}
\includegraphics[width=0.98\linewidth]{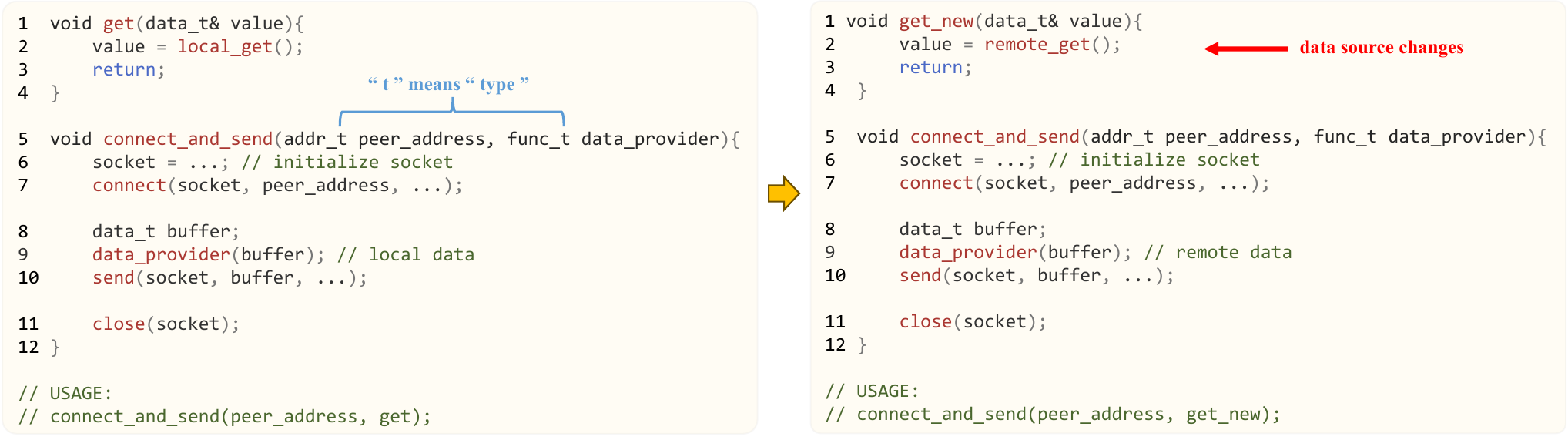}
}

\quad
\vspace{-0.3cm}

\subfigure[Simulation implementation of local data fetch and direct update to remote data fetch]{
\label{fig:callback_bug_code_simulation}
\includegraphics[width=0.98\linewidth]{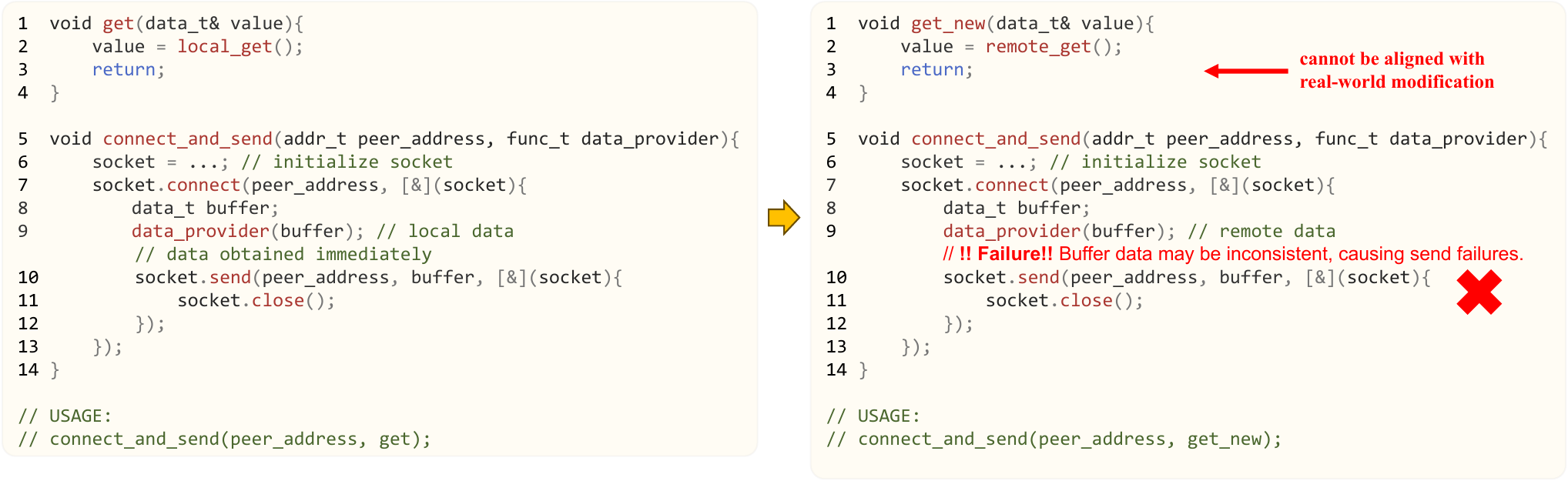}
}

\vspace{-0.3cm}

\caption{Real-world and callback-based simulation pseudocode of data fetching and sending under data source extension.}
\label{fig:callback_bug_code}
\vspace{-0.3cm}
\end{figure*}

\begin{figure}[htbp]
    \centering
    \vspace{-0.3cm}
    \includegraphics[width=0.45\textwidth]{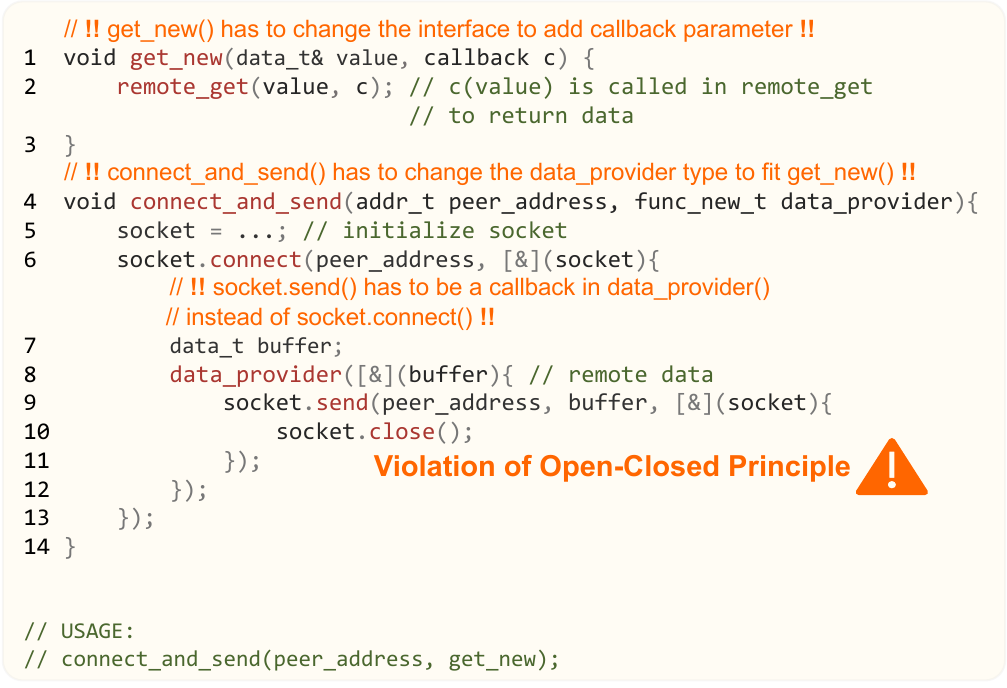}
    \caption{Simulation implementation of update to remote data fetch.}
    \vspace{-0.6cm}
    \label{fig:callback-bug-code-violation}
\end{figure}


\subsection{Why are Callbacks Used}

Most existing network simulators are based on DES, in which events are scheduled
and processed in chronological order \cite{DES-intro}. In DES, an event combines a
future operation with a timestamp indicating when it should execute, and once
triggered, it invokes a handler to carry out the operation and trigger subsequent logic. This execution model naturally calls for callbacks, which serve as the handlers that the scheduler invokes upon event dispatch, linking one event's completion to the next operation \cite{ns3-2010}.

Callbacks also reduce direct inter-module dependencies compared to earlier tightly
coupled designs, where simulation models directly invoked each other's methods
\cite{2008-ns3, ns3-2010}. By mediating interactions through callbacks, modules
can be added, updated, and compiled independently, improving extensibility,
modularity, and compilation efficiency. These properties make callbacks the de
facto standard in DES-based network simulators. However, as network functions
grow in complexity, their limitations become increasingly pronounced.


\subsection{Problems of Callbacks}
\label{subsection:callback_problem}

Despite their widespread use and apparent suitability for modeling network events,
callbacks exhibit substantial limitations as code complexity grows, manifesting at
two levels: structural violations of design principles, and day-to-day development
and maintenance burdens.

At the structural level, callbacks deviate from their original intent and violate
the Open/Closed Principle \cite{open-closed-principle} once blocking operations
are introduced. We provide a basic networking scenario as an example in
Figure \ref{fig:callback_bug} and its corresponding implementation in
Figure \ref{fig:callback_bug_code}. Procedure A, a typical network function
\textit{connect\_and\_send()}, first fetches required data and then sends it;
Procedure B is a \textit{get()} function that returns the data. In the initial
scenario (Figure \ref{fig:callback_bug_1}), the data source responds immediately,
so Procedure B returns at once, enabling Procedure A to obtain the data and
proceed. As new requirements emerge, however, the data source may need to support
delayed retrieval (e.g., remote cloud storage), which introduces blocking
(Figure \ref{fig:callback_bug_2}). Under the Open/Closed Principle, the interface
between Procedures A and B should remain unchanged; yet once Procedure B becomes
blocking, the original callback structure provides no mechanism to deliver the
result back to Procedure A upon completion, preventing the subsequent send from
executing correctly. This breaks the Open/Closed Principle, as extending behavior
requires modifying existing interfaces.

To further illustrate this issue, Figure \ref{fig:callback_bug_code} compares
concrete implementations in real systems and in simulation. In real systems
(Figure \ref{fig:callback_bug_code_realworld}), blocking behavior can be
introduced without changing function interfaces because the operating system
manages blocking events transparently. In DES, by contrast, blocking must be
modeled explicitly through callbacks. As shown in
Figure \ref{fig:callback_bug_code_simulation}, the same requirement change cannot
be addressed by modifying Procedure B alone. Instead, it necessitates changing
the interface of Procedure B to accept an additional callback and updating
Procedure A accordingly (Figure \ref{fig:callback-bug-code-violation}). This
change increases callback nesting and structural complexity, breaks interface
stability, increases inter-module coupling, and violates the Open/Closed Principle.
As the codebase grows, such callback dependencies become increasingly tangled; the
two procedures shown here form a minimal example, and real network simulators
typically involve much more complex interaction patterns.

\begin{figure}[htbp] 
\vspace{-0.8cm}
\centering

\subfigure[Real-world procedure of data fetch and send]{
\label{fig:callback_problem_a}
\includegraphics[width=0.94\linewidth]{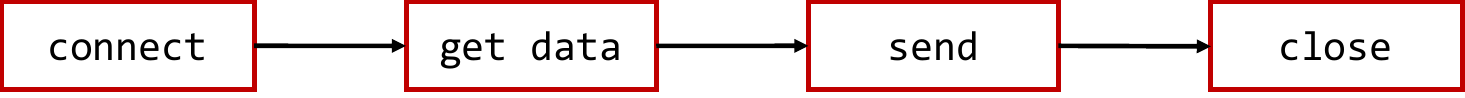}
}
\quad
\vspace{-0.2cm}

\hspace{-0.12cm}
\subfigure[Callback implementation procedure for data fetch and send]{
\label{fig:callback_problem_b}
\includegraphics[width=0.9\linewidth]{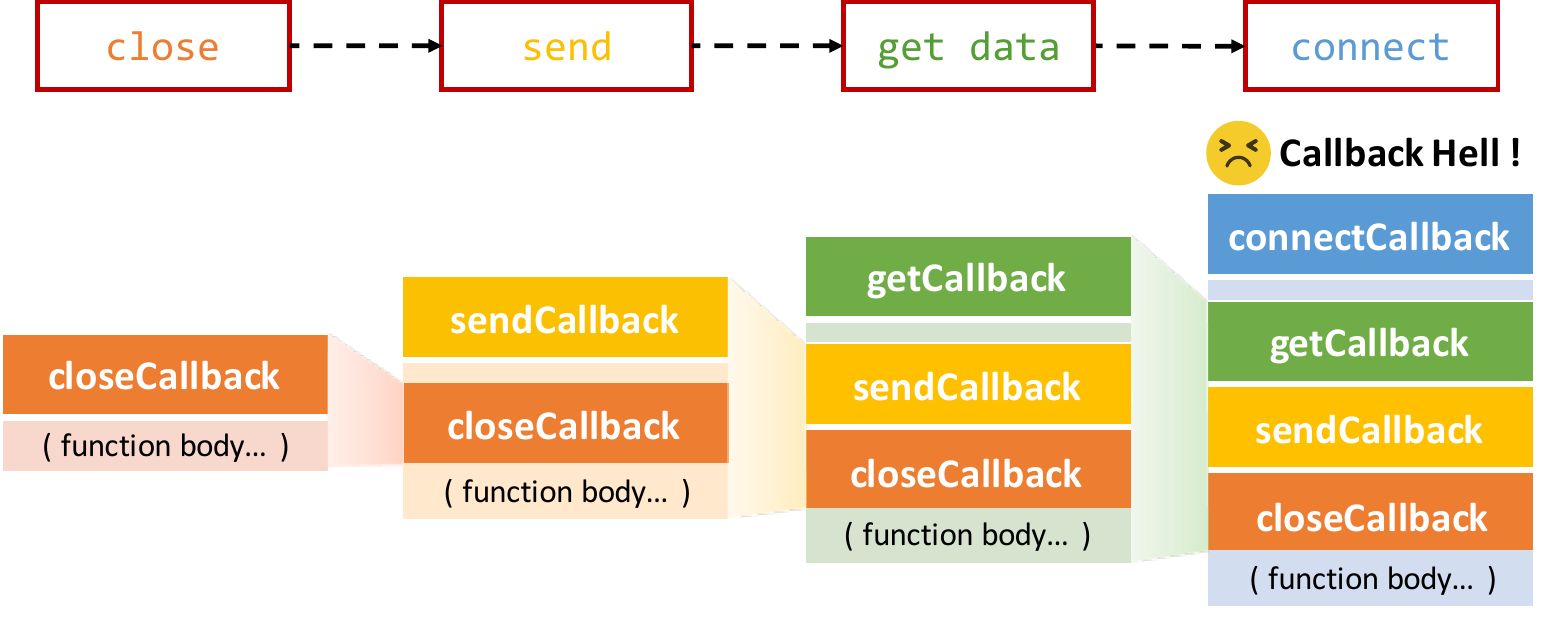}
}
\caption{``Callback Hell" problem in existing network simulation.}
\label{fig:callback_problem}
\vspace{-0.8cm}
\end{figure}

At the development and maintenance level, nested callbacks impose substantial
burdens throughout the development lifecycle. The use of deeply nested callbacks
makes code structure chaotic and difficult to manage, a problem commonly referred
to as ``callback hell'' \cite{depfast, stack-ripping, edwards2009callbackhell}.
In network simulation this problem is particularly severe, given the need to manage
numerous complex network states such as packet loss handling, timeout
retransmissions, and multi-node communication \cite{callback-hell-3}. As shown in
Figure \ref{fig:callback_bug_code_simulation}, even a relatively simple function
already exhibits increased nesting in its simulation implementation; in more complex
scenarios the structure becomes far harder to develop and maintain. For instance,
MPI \cite{dongarra1995MPI}, a typical protocol for parallel-system communication,
is implemented with approximately 10K LOC and more than 15 levels of nesting depth
in SST/Macro \cite{2015-sst, 2016-sstbooksim}. As illustrated in
Figure \ref{fig:callback_problem}, inverted control flow forces developers to
implement later callbacks first to satisfy earlier interfaces, imposing additional
cognitive overhead that compounds the structural complexity.

\begin{figure}[t]
    \centering
    \vspace{-0.9cm}
    \includegraphics[width=0.35\textwidth]{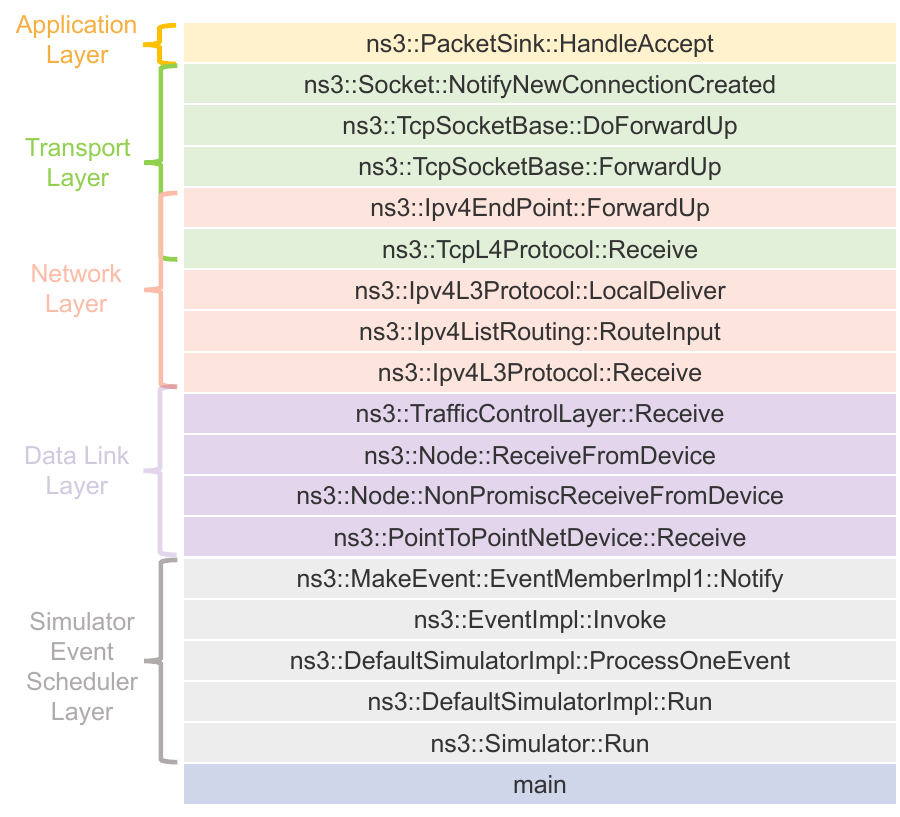}
    \caption{Debugging stack of handling received packets on the application layer under callback-based simulation.}
    \vspace{-0.5cm}
    \label{fig:callback-debug}
\end{figure}

The nested structure induced by callbacks also complicates debugging.
Figure \ref{fig:callback-debug} shows a debugging stack for a basic
application-level function. When a breakpoint is set at the application-layer
handler for received packets, the stack reveals multiple network layers traversed
by the packet, with call frames from different layers interleaved (e.g., transport
and network layers). Extracting useful information from such a deep and entangled
stack is difficult, and the application-level context that developers actually need
is obscured by largely irrelevant low-level details.

Both the structural violations and the development burdens described above share a
common root: \textbf{callbacks do not naturally preserve the sequential structure of blocking network operations in DES.}
In real-world programs, the OS transparently suspends and resumes blocking
operations. In DES, however, a single callback can represent only one point in a
blocking operation, its start or completion, but cannot capture the full operation
as a coherent whole. As a result, callbacks provide only a fragmented view of
blocking behavior, forcing developers to manually decompose sequential logic into
scattered callback fragments. Our survey of over 100 PhD students, engineers, and
researchers in networking confirms these findings: participants report heavy
development workloads and poor maintainability in callback-based simulation.

\subsection{What are Coroutines}
\label{subsubsection:coroutine_basics}

The core idea of coroutines is to allow a function to suspend execution, yield
control, and later resume from the suspension point, thereby enabling asynchronous
operations to be expressed as sequential code. This mechanism is fully supported
in C++20 \cite{C++coroutine-2}, the language upon which most network simulators
are built \cite{2008-ns3, opnet, omnet++, glomosim}, and is applied in complex environments such as IoT and embedded systems
\cite{belson2019coroutin-embedded-survey}.

It is important to distinguish coroutines from threads. Coroutines are
cooperatively scheduled in user space and rely on an event loop provided by the
runtime or libraries, with relatively low memory overhead of only tens to hundreds
of bytes per instance \cite{C++coroutine-1, C++coroutine-2} and low
context-switch overhead. Threads, in contrast, are preemptively scheduled by the
kernel and typically incur a much heavier burden, with individual sizes reaching
several megabytes \cite{Nishanov2018Fibers}, and their overhead grows significantly
as the number of concurrent tasks increases. More critically for DES, threads
provide no mechanism for explicitly transferring control to a designated routine,
since scheduling is fundamentally non-deterministic. This incompatibility with
DES's requirement for precise event ordering makes threads unsuitable as the
foundation for a DES programming paradigm.

Since threads are unsuitable for DES, one might consider fibers as a lighter alternative that also supports cooperative scheduling. However, fibers introduce their own structural incompatibility with DES. Fibers utilize a dedicated scheduler and maintain a full, pre-allocated call stack of often several kilobytes \cite{Nishanov2018Fibers}, which introduces architectural redundancy when integrated with the native event loop of DES. In contrast, coroutines are stackless and managed directly via language constructs like \textit{co\_await}, storing only minimal state. Consequently, coroutines exhibit significantly lower memory usage and context-switching overhead, approximately $20\times$ lower than fibers \cite{Nishanov2018Fibers}, making them far better suited to the resource constraints and scalability demands of network simulation. Indeed, OMNeT++'s stackful \textit{activity()} model, which relies on platform-specific fiber APIs, is officially deprecated in favor of callbacks due to slow context switching and debugger incompatibility \cite{omnet++}, demonstrating that naive fiber adoption does not suffice for scalable DES.

These properties, namely cooperative scheduling, minimal state overhead, and explicit control transfer, make coroutines a natural fit for DES, motivating the paradigm we describe next.

\subsection{A New Paradigm}

Building on the coroutine mechanism, we propose CoDES, a programming paradigm
that expresses each blocking network operation as a single coroutine function,
suspending at blocking points and resuming upon event completion, thereby
restoring the sequential workflow that callbacks fragment. Unlike
resource-constrained approaches that abandon local coroutine state in favor of
global variables \cite{protothreads}, CoDES preserves the necessary execution
state per coroutine, enabling correct handling of concurrent network operations
without manual state management.

\begin{figure}[hbp]
    \centering
    \vspace{-0.2cm}
    \includegraphics[width=0.48\textwidth]{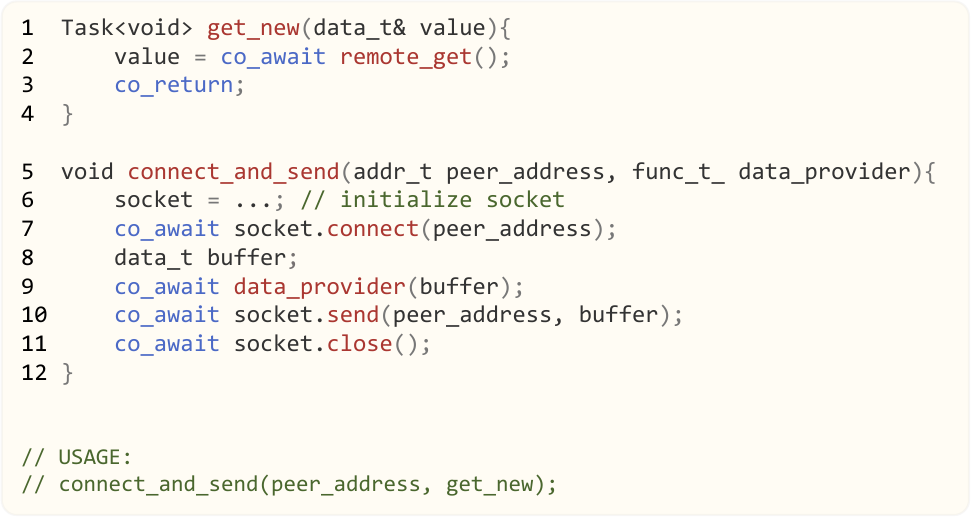}
    \caption{Simulation implementation of CoDES-based data fetch and send example.}
    \vspace{-0.2cm}
    \label{fig:codes-example}
\end{figure}

\begin{figure}[htbp]
    \centering
    \vspace{-0.1cm}
    \includegraphics[width=0.4\textwidth]{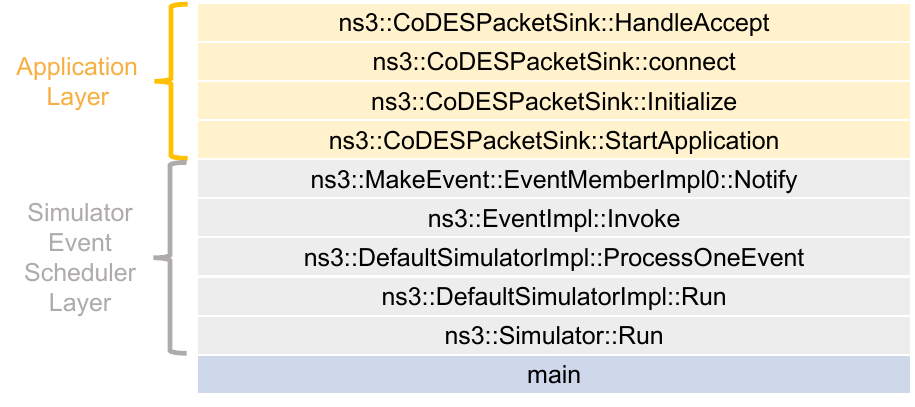}
    \caption{Debugging stack of handling received packets on the application layer under coroutine-based simulation.}
    \vspace{-0.2cm}
    \label{fig:coroutine-debug}
\end{figure}

CoDES directly addresses the two levels of limitations identified in
Section \ref{subsection:callback_problem}.
At the structural level, CoDES restores extensibility for module upgrades and
feature additions. Using CoDES, we rewrite the data-fetching and sending simulation
program (Figure \ref{fig:codes-example}). The resulting code remains natural and
sequential: irrespective of whether data retrieval involves blocking operations,
the fetched data is returned in the usual manner before being sent, without
requiring interface modifications, directly resolving the cascading interface
changes described in Section \ref{subsection:callback_problem}. CoDES also
substantially reduces structural complexity, directly addressing the deep nesting
and inverted development order that make callback-based implementations such as
MPI in SST/Macro difficult to develop and maintain, as quantitatively evaluated
in Section \ref{sec:evaluation}.
At the development and maintenance level, CoDES yields a markedly clearer debugging
experience. Figure \ref{fig:coroutine-debug} shows the debugging stack for the
same application-level function under CoDES-based simulation. In contrast to the
convoluted stack trace in Figure \ref{fig:callback-debug}, CoDES produces an
application-oriented view that avoids irrelevant low-level details and preserves
the application-layer context at the breakpoint, providing more actionable
debugging information. As confirmed by our survey,
85.7\% of respondents believe that the CoDES-based implementation reduces the
development burden compared with the callback-based counterpart.

\section{C\lowercase{o}DES Overview}
\label{sec:overview}


CoDES enables every blocking network operation to be expressed as a
sequential coroutine function, with suspension and resumption handled
transparently by the framework. Achieving this in DES requires bridging
two independent control flows: coroutines follow program logic, whereas
the event scheduler drives execution. CoDES bridges them by translating
each coroutine suspension point into a schedulable DES event, so that
sequential coroutine code becomes executable within the discrete-event model.


Users interact with CoDES by encapsulating each network operation as a CoDES
operation, with all state management handled automatically by the framework.
\textit{co\_await} waits for a blocking operation to complete and obtains the
result.
\textit{co\_yield} actively suspends the current coroutine and yields control,
to be resumed when an external condition is met. 
\textit{co\_return} terminates the current operation and delivers
the final result to its caller.


The framework consists of two components, namely state maintenance and CoDES
operations. State maintenance preserves and restores execution state across
suspension and resumption. A CoDES operation handles the complete lifecycle of
a network operation, whether blocking or non-blocking, including creation,
execution, suspension, and termination.


Integrating coroutines into an existing DES simulator surfaces four non-trivial
challenges that do not arise in purely callback-based designs.
\textbf{1) Correctness:} concurrent coroutines may hold cyclic references to
shared simulation resources, which must be broken to prevent memory
leaks and deadlocks.
\textbf{2) Functionality:} network algorithms frequently require composing results
across multiple concurrent operations, such as anycast
\cite{anycast-1, anycast-2} and load balancing
\cite{load-balancing-1, load-balancing-2}, demanding
operation-level composition primitives beyond what a single coroutine naturally
provides.
\textbf{3) Compatibility:} CoDES must interoperate with existing callback-based
simulators without a full rewrite, demanding a provably correct bidirectional
translation between the two paradigms.
\textbf{4) Overhead:} thousands of concurrent coroutines may accumulate
significant simulation slowdown even with modest per-coroutine cost, making
overhead minimization a first-class concern.
We address all these challenges in Section \ref{sec:design}.

\section{Why Callbacks Cannot Naturally Express Blocking Operations}
\label{sec:theory}

Before elaborating on the system-level design, we provide a
theoretical grounding for the central claim of
Section \ref{subsection:callback_problem}, drawing on the operational
semantics of coroutines \cite{moura2009coroutinetheory} and
continuation-passing style (CPS) transformation theory
\cite{reynolds1993discoveries, appel1992compiling}, the formal
basis of the callback style.


\subsection{Definitions of Callback and Coroutine}

We model the simulation runtime state as a tuple $(\Sigma, \Delta)$,
where $\Sigma$ represents the global state (e.g., the simulation event
queue) and $\Delta$ represents the local execution context
(e.g., local variables).

\begin{definition}[Callback]
\label{def:callback}
A callback-based event handler is a standard subroutine. Let
$\mathcal{F}_{cb}$ be the set of callback functions. An execution of a
callback $f \in \mathcal{F}_{cb}$ is modeled as the transition:
\begin{equation}
\langle f,\,\Sigma,\,\Delta_{in} \rangle \xrightarrow{\,\mathrm{call}\,}
\langle \Sigma',\,\bot \rangle
\end{equation}
Upon completion, the local context $\Delta_{in}$ is destroyed
($\bot$). Consequently, representing a blocking operation that resumes
with intermediate state intact requires externalizing that state
into $\Sigma$.
\end{definition}

\begin{definition}[Coroutine]
\label{def:coroutine}
Following \cite{moura2009coroutinetheory}, a coroutine is a state
machine equipped with \textbf{yield} and \textbf{resume} primitives.
The \textbf{yield} operation suspends execution while preserving the
local context:
\begin{equation}
\langle c,\,\Sigma,\,\Delta \rangle \xrightarrow{\,\mathrm{yield}\,}
\langle c',\,\Sigma',\,\Delta \rangle_{\mathrm{susp}}
\end{equation}
The \textbf{resume} operation restores the suspended state and
continues execution, with $\Delta$ intact throughout.
Crucially, $\Delta$ is preserved across the yield--resume cycle,
enabling coroutines to retain intermediate state without
externalization.
\end{definition}


\subsection{The Structural Cost of Blocking in Callbacks}

The callback style is the concrete realization of
CPS in event-driven systems: each
callback encodes exactly one continuation (the ``what to do next''
upon event completion). A foundational result in programming language
theory is that transforming a conventional sequential program into CPS requires a
\emph{global} program restructuring: every function along the call
chain must be rewritten to accept an explicit continuation argument
\cite{reynolds1993discoveries, appel1992compiling}. In conventional
sequential code, a function may block and return its result without
any change to surrounding code; in CPS, the same blocking point
forces every caller along the chain to accept and pass an
explicit callback.

In DES, each callback is invoked exactly once upon event dispatch,
after which its local context is destroyed
(Definition \ref{def:callback}). This matches precisely the CPS
model, so the globality result applies directly:

\begin{corollary}[Structural Cost of Blocking]
\label{cor:blocking-cost}
Let $C$ be a $k$-phase blocking operation ($k \geq 1$ suspension
points) carrying live local variables $V \subseteq \Delta$ across
phases. Its callback representation requires:
{\small
\begin{equation}
\mathcal{T}(C) \;\Rightarrow\;
\bigl\{\{S_i\}_{i=1}^{k},\;
f_0(\Sigma)\!\mapsto\! S_1,\;
f_i(\Sigma, S_i)\!\mapsto\! S_{i+1},\;
f_k(\Sigma, S_k)\bigr\}
\end{equation}
}
where $\{S_1,\ldots,S_k\}$ are $k$ new global structures externalizing
$V$ (necessitated by $\Delta \to \bot$ in Definition \ref{def:callback}),
and $\{f_0,\ldots,f_k\}$ are $k{+}1$ disjoint callback fragments.
The size of $\mathcal{T}(C)$ grows linearly with $k$: any change
to the number or nature of suspension points forces modifications
throughout the call graph.
\end{corollary}

In contrast, a coroutine represents the same $k$-phase operation as a
single coherent function: $\Delta$ is preserved automatically across all
$k$ yield points (Definition \ref{def:coroutine}), requiring no new
global structures and no fragmentation. Moura and Ierusalimschy
establish that coroutines subsume callbacks (any callback is embeddable
into a coroutine through a local wrapping
\cite{moura2009coroutinetheory}):
{\small
\begin{equation}
\mathrm{Wrap}(f) \;=\; \mathbf{coroutine}\;\{\; f();\;\mathbf{return};\;\}
\end{equation}
}
This wrapping is purely local: it modifies no internals of $f$,
introduces no new global state into $\Sigma$, and leaves the
event-scheduling structure unchanged. The reverse direction requires
the full CPS transformation of Corollary \ref{cor:blocking-cost},
an asymmetry consistent with Felleisen's macro-eliminability criterion
\cite{felleisen1991expressive}.

The cascading interface modifications in
Figure \ref{fig:callback_bug_code} are a direct instance of this
global restructuring: one new blocking phase in \textit{get()} forces
\textit{connect\_and\_send()} to be rewritten, matching exactly the
$k{=}1$ case of Corollary \ref{cor:blocking-cost}. The
``stack ripping'' phenomenon \cite{stack-ripping} is the
developer-visible consequence as this cost accumulates across deeply
nested call graphs. CoDES resolves this by substituting coroutines
for callbacks, restoring sequential expression of blocking operations
without manual state management.

\section{C\lowercase{o}DES Design}
\label{sec:design}


CoDES comprises two core components, namely state maintenance and CoDES
operations. State maintenance stores and manages the execution states of network
operations, while CoDES operations define the lifecycle and composition interfaces
through which those operations are created, awaited, and chained. The following
sections detail these two components and demonstrate their compatibility with
existing callback-based network simulators. CoDES is protocol-agnostic and can
be instantiated over any network stack.

\subsection{State Maintenance}

Reliable state preservation is a prerequisite for coroutine-based execution in
DES. The coroutine frame structure is the core of state maintenance in CoDES,
responsible for storing all execution state pertaining to a network operation,
including local variables, on-complete handlers, and references to resources such
as network connections. When a coroutine awaits a network operation, CoDES
suspends the coroutine and registers a completion handler on the underlying
scheduled event. When that event is triggered, the handler resumes the
coroutine, restoring all state from the frame structure to continue execution
exactly where it left off.

Because the frame structure holds references to shared simulation resources,
cyclic ownership can arise when the same resource is referenced by multiple
frame structures, leading to memory leaks \cite{2022-memoryleak} or deadlocks.
To address this correctness challenge \textbf{(this solution corresponds to
challenge 1)}, we transform the resource reference graph
\cite{klabnik2023cyclic-reference} into a directed acyclic graph, assigning
exclusive ownership \cite{josuttis2012c++} of each resource to the single event
currently holding it while all others maintain non-owning references. Cycle
prevention is structural rather than algorithmic: ownership flows strictly from
the scheduler to the operation via the completion callback, with no back-edges,
so cycles cannot form by construction. This is feasible because single-process
DES simulators \cite{2008-ns3,opnet,omnet++} execute events sequentially,
ensuring that only one event accesses a resource at any given time.

A second correctness concern arises when multiple callers hold references to the
same frame structure concurrently, risking premature or duplicate release. We
therefore enforce that exactly one caller holds exclusive ownership
\cite{josuttis2012c++} of each frame structure at any time, controlling its
lifecycle, while other callers access its state data in a read-only capacity
\textbf{(this solution corresponds to challenge 1)}. To guarantee the frame
structure is released exactly once, we introduce a reference counter that tracks
the number of active holders and triggers destruction only when the last holder
finishes. This design also ensures that all resume and terminate operations are
idempotent: once a frame has completed, any subsequent attempt to resume or
destroy it is a safe no-op, ruling out double-resume and dangling access
\textbf{(this solution corresponds to challenge 1)}.

\begin{figure*}[htbp]
    \centering
    \vspace{-0.6cm}
    \includegraphics[width=0.95\textwidth]{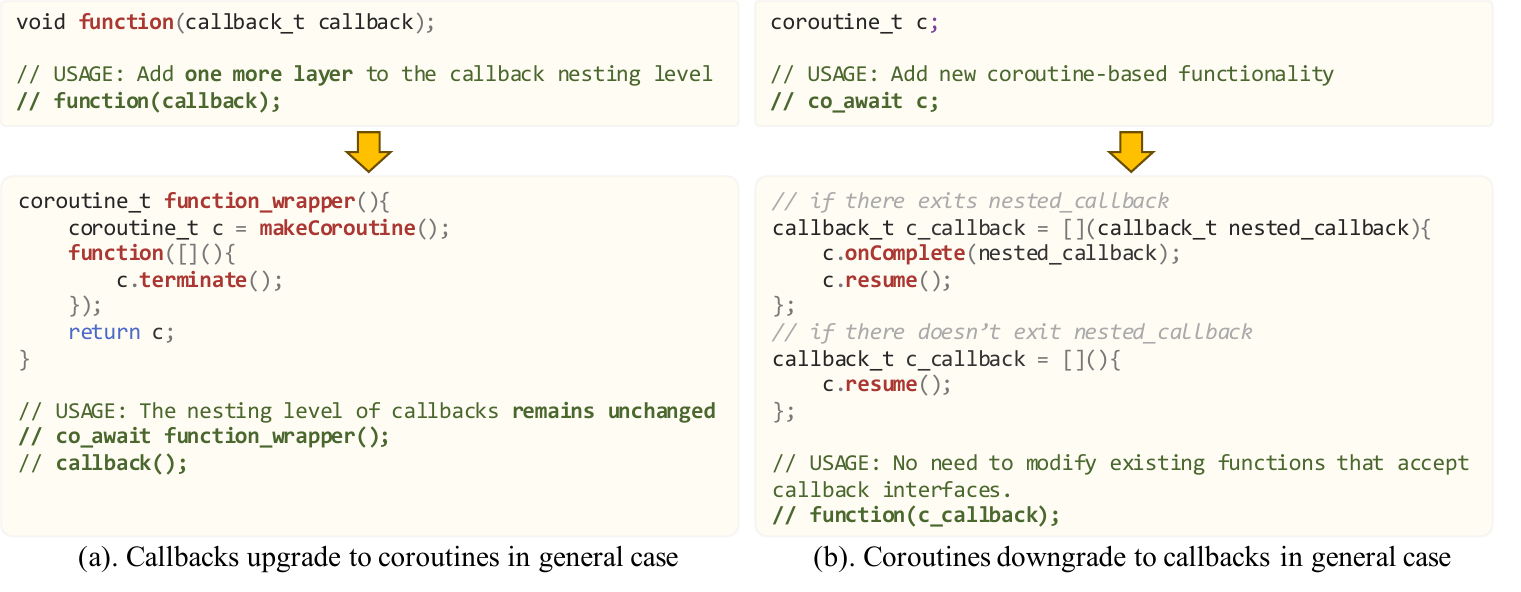}
    \vspace{-0.3cm}
    \caption{Adapting CoDES to existing callback-based network simulation.}
    \label{fig:transformation}
    \vspace{-0.5cm}
\end{figure*}

With correctness ensured, the remaining concern for state maintenance is
overhead. CoDES employs C++20 stackless coroutines \cite{C++coroutine-2} to
address this challenge \textbf{(this solution corresponds to challenge 4)}. In
DES, execution is driven by a single-threaded event scheduler, which means each
coroutine runs to its next suspension point without preemption. This property
makes stackless coroutines particularly well suited: rather than maintaining a
full per-coroutine call stack, CoDES saves only the minimal execution state at
each suspension point, namely the local variables and the program counter. The
resulting memory footprint is tens to hundreds of bytes per coroutine, and
context switching reduces to a simple state restore \cite{C++coroutine-2}.
Although managing this minimal state correctly across arbitrary suspension points
requires careful implementation, the complexity is fully encapsulated within
CoDES and remains invisible to developers.


\subsection{C\lowercase{o}DES Operation}
\label{sec:operation}

Building on the state maintenance mechanism, CoDES operations provide the
developer-facing interface for expressing network operations as coroutines.
A CoDES operation encapsulates any network operation as a coroutine, allowing
execution to be paused at blocking points and resumed upon event completion. Its
execution state is managed by the coroutine frame structure and coroutine handle,
which together track the current suspension point and lifecycle. Execution logic
is expressed directly as sequential code using \textit{co\_await} and
\textit{co\_yield}. Return values and exceptions are propagated through the frame
structure upon completion, and registered on-complete handlers are invoked
automatically.

As a concrete realization of CoDES operations at the transport layer,
\textit{CoroutineSocket} exposes each socket primitive (accept, connect, send,
receive) as a \textit{CoroutineOperation}, allowing upper-layer protocol logic
to await socket events directly without managing callbacks. Internally,
\textit{CoroutineSocket} maintains four pending operation queues corresponding
to each socket event type. When a socket event is triggered, the
corresponding queued operation is resumed, completing the awaiting coroutine and
returning the result to the caller.

With individual operations expressible and awaitable, the remaining functionality
challenge concerns how operations can be composed to express more complex network
behaviors. Existing network algorithms frequently require coordination across
multiple concurrent operations. We therefore introduce \textit{OperationCalculation}
as a higher-level composition primitive
\textbf{(this solution corresponds to challenge 2)}. In typical network scenarios
such as anycast \cite{anycast-1, anycast-2} and load
balancing \cite{load-balancing-1, load-balancing-2}, one sender
issues requests to multiple receivers and can proceed once the first response
arrives. \textit{OperationCalculation} supports operations of the form
\(\,op_1 \parallel op_2 \parallel op_3\,\), unifying the result types of the
concurrent operations. Once any operation completes, the remaining ones are
terminated via cooperative cancellation: the first completing operation sets a
termination flag on each sibling, causing them to unwind at their next suspension
point without requiring explicit removal of events from the scheduler queue.

Beyond parallel composition, CoDES also supports sequential transformation of
operation results through \textit{ChainOperation}, inspired by lazy evaluation
\cite{launchbury1993lazyevaluation} and chaining \cite{martin2009chaining} from
functional programming \cite{hughes1989functionalprogramming1}
\textbf{(this solution corresponds to challenge 2)}.
\textit{ChainOperation} condenses multi-layered nested operations into a single
operation and eliminates intermediate variables, reducing structural complexity.
In the CoDES implementation, \textit{ChainOperation} is realized through the
\textit{then()} method on \textit{CoroutineOperation}, which chains a
transformation function onto an existing operation and returns a new operation
that automatically awaits the original before applying the function to its result.
The lazy evaluation strategy of \textit{ChainOperation} decouples the dependency
between communication information and the algorithm itself, reducing unnecessary
coupling. Developers can pass a \textit{ChainOperation} to subsequent layers
without concern for when or how the underlying communication completes, enabling
more focused development and debugging of network algorithms.


\subsection{C\lowercase{o}DES Compatibility}
\label{sec:compatibility}

Existing network simulation environments are based on callbacks. CoDES supports
bidirectional translation between coroutines and callbacks, allowing it to reuse
existing callback-driven functionality while providing coroutine-style benefits.
General scenarios are outlined below. Multiple network functions are
integrated into ns-3 as detailed in Section \ref{sec:implementation}.

Figure \ref{fig:transformation}(a) illustrates the general scenario of upgrading
callbacks to coroutines \textbf{(this solution corresponds to challenge 3)}.
Since any callback can be transformed into a coroutine body with only an exit
point, as stated in Section \ref{sec:theory}, we wrap the original function and
replace the callback with such a coroutine. When execution reaches the callback
site, control yields to the caller, which then invokes the callback in a
sequential workflow, without increasing nesting depth.

The complementary scenario arises when new CoDES-based features need to
interoperate with existing callback-based interfaces.
Figure \ref{fig:transformation}(b) depicts this general downgrade scenario
\textbf{(this solution corresponds to challenge 3)}. A full rewrite of existing
interfaces is not required. Instead, the new feature is encapsulated within a
callback wrapper and registered with the existing callback-based function, which
then triggers the CoDES-based feature as expected. As stated in
Section \ref{sec:theory}, one coroutine can be downgraded to several callbacks
when necessary. Both translation directions preserve event ordering and
exception semantics.

\section{Building Network Functions with C\lowercase{o}DES}
\label{sec:implementation}

To demonstrate the effectiveness of CoDES, we implement three representative
network functions in ns-3 \cite{ns3-2010, 2008-ns3}: MPI \cite{walker1996mpi,
dongarra1995MPI}, HPCC \cite{2019-sigcomm-HPCC}, and RIP \cite{stark1999rip}. These cases span the transport, congestion control, and routing
layers respectively, and each represents a class of state management complexity
that is fundamentally difficult to express with callbacks but maps naturally onto
coroutines.

MPI is a communication standard for
distributed parallel systems that coordinates
operations across processes under strict execution dependencies \cite{walker1996mpi, dongarra1995MPI}. With CoDES,
inter-process synchronization becomes straightforward to express. In Ring
AllReduce, for example, each step requires every node to issue a send and a
receive simultaneously and to proceed only after both complete. Developers initiate
both operations and use \textit{co\_await} to wait for their joint completion,
without manually tracking which callback has fired or managing intermediate
synchronization state.

HPCC is a congestion control protocol for 
RDMA networks that requires fine-grained timed queue state management driven by
packet arrivals and timer events \cite{2019-sigcomm-HPCC}. With CoDES, time-sensitive state transitions
can be expressed and cancelled cleanly. In PFC pause recovery, for example,
monitoring both a timer expiry and a possible refresh from a new PFC frame is
naturally expressed by \textit{co\_await}-ing a timed operation and
terminating it explicitly on refresh, avoiding the ABA problem \cite{dechev2010ABA} that
arises when callbacks poll a shared state flag.

RIP is a distance-vector routing protocol
that requires coordinating two interleaved asynchronous triggers, namely periodic
timer-driven updates and event-driven route message receptions, both modifying
shared routing table state \cite{stark1999rip}. With CoDES, both triggers are expressed as
\textit{co\_await} operations within the same sequential function, keeping all
routing table update logic co-located rather than scattered across independent
callback handlers, and making state consistency easier to reason about and
maintain.

\section{Evaluation}
\label{sec:evaluation}
In this section, we address three questions: 1) How much development effort can CoDES save? 2) Does CoDES benefit AI-assisted code generation? 3) What overhead does CoDES introduce in accuracy, execution time, and memory? We answer these by comparing CoDES-based implementations of MPI, HPCC, and RIP against their callback-based counterparts.

\begin{figure*}[htbp] 
\vspace{-0.9cm}
\centering

\subfigure[Token Consumption]{
\label{fig:token_consumption}
\includegraphics[width=0.3\textwidth]{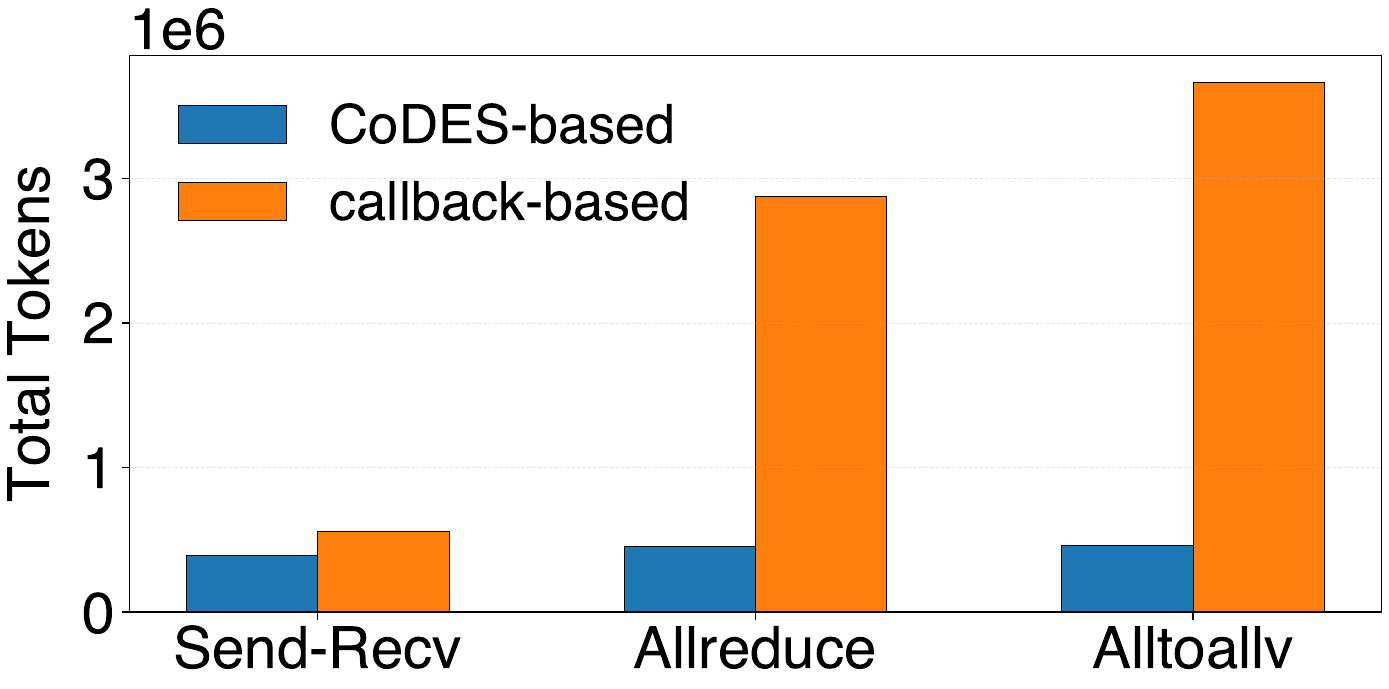}
}
\hfill
\subfigure[Generation Time]{
\label{fig:generation_time}
\includegraphics[width=0.3\textwidth]{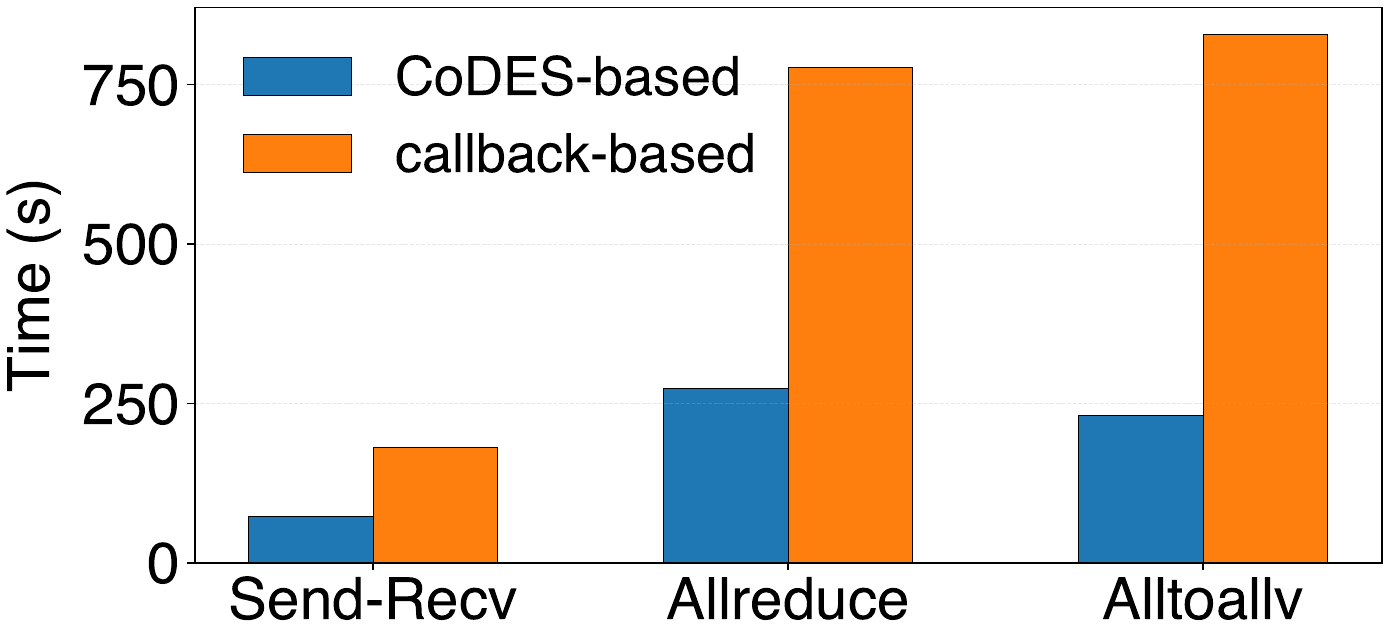}
}
\hfill
\subfigure[Number of Error Corrections]{
\label{fig:error_corrections}
\includegraphics[width=0.3\textwidth]{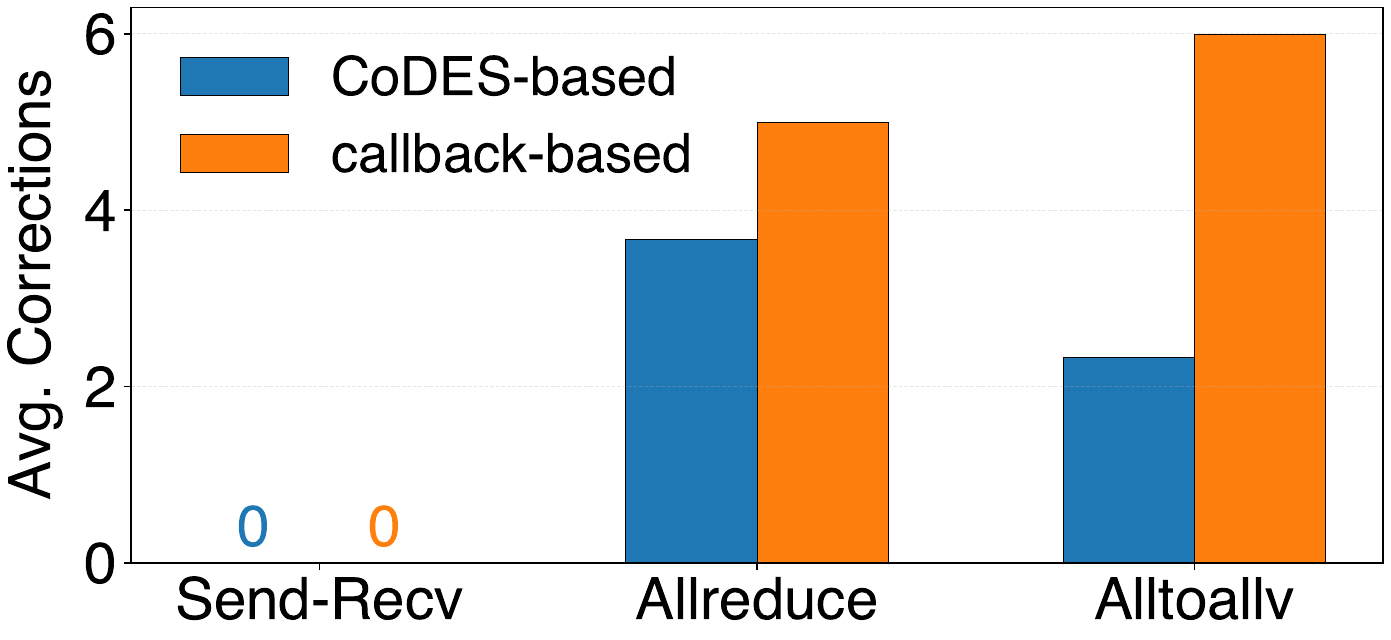}
}

\vspace{-0.3cm}

\caption{Cost and quality metrics for AI-assisted code generation under callback-based and CoDES-based paradigms.}
\label{fig:ai-evaluation}
\vspace{-0.6cm}
\end{figure*}

\subsection{System Setup}
We implement CoDES-based MPI, HPCC, and RIP in ns-3 \cite{ns3-2010, 2008-ns3, hayamizu2019ceforesim}, one of the most widely adopted discrete-event network simulators in academic research (used by 90.5\% of surveyed researchers), alongside their callback-based counterparts as baselines. The MPI baseline is reproduced in ns-3 based on SST/Macro \cite{2015-sst, 2016-sstbooksim}, a production-grade callback-based simulator, while the HPCC and RIP baselines are adopted from existing ns-3 implementations.

Experiments run on a server equipped with an AMD EPYC 7R13 48-core processor and 251 GiB of memory. The benchmarks for MPI include various MPI functions and typical real-world traces from HPCG \cite{heroux2013HPCG}, LULESH \cite{karlin2013lulesh}, and HILO \cite{2016-sstbooksim}.  For HPCC and RIP, we use the official ns-3 test suites for verification via system and unit tests. 

\subsection{Code Volume \& Structure Complexity}
In our experience, master's students with no prior simulation background require
approximately 2 person-months to learn C++20 coroutines and 4 person-months to
implement CoDES-based MPI, compared with 3 person-months of learning and
15 person-months of development for the callback-based counterpart. To address the first question more rigorously, we compare the LOC, nesting depth, and cyclomatic complexity \cite{mccabe1976cyclomaticcomplexity} of callback-based versus CoDES-based implementations.

\begin{figure}[htbp]
    \centering
    \vspace{-0.4cm}
    \includegraphics[width=0.4\textwidth]{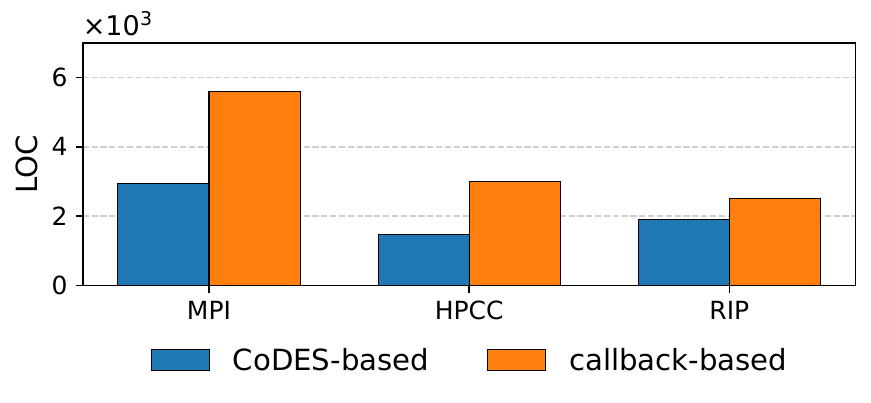}
    \vspace{-0.2cm}
    \caption{LOC comparison between CoDES-based and callback-based implementations.}
    \vspace{-0.3cm}
    \label{fig:LOC}
\end{figure}

\subsubsection{Lines of Code (LOC)}
As shown in Figure \ref{fig:LOC}, the CoDES-based implementations reduce the LOC by 24.7\%-51.4\% compared to the callback-based implementations. As the complexity of states and operations increases, e.g., from RIP to MPI, the reduction in code volume achieved by CoDES becomes more significant.

\begin{figure}[htbp]
    \centering
    \vspace{-0.3cm}
    \includegraphics[width=0.4\textwidth]{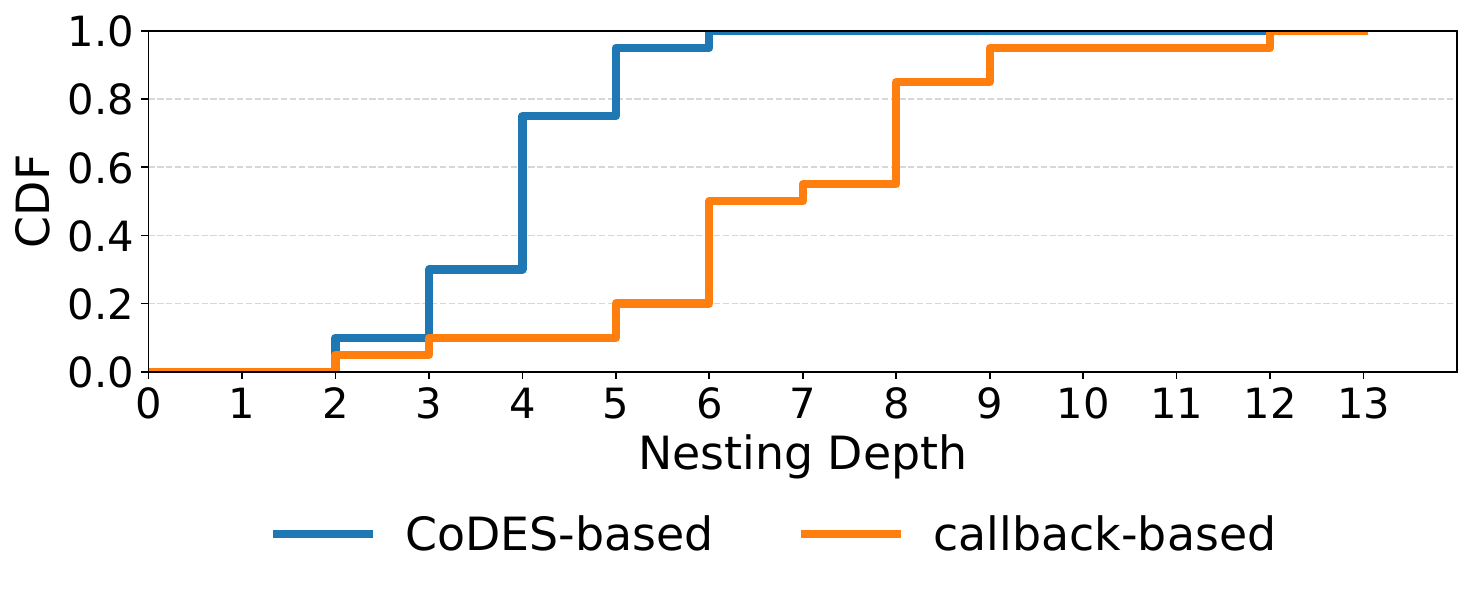}
    \vspace{-0.2cm}
    \caption{Nesting depth comparison of 20 classical MPI functions in CoDES-based and callback-based implementations.}
    \vspace{-0.3cm}
    \label{fig:nesting_depth}
\end{figure}

\begin{figure}[htbp]
    \centering
    \vspace{-0.2cm}
    \includegraphics[width=0.48\textwidth]{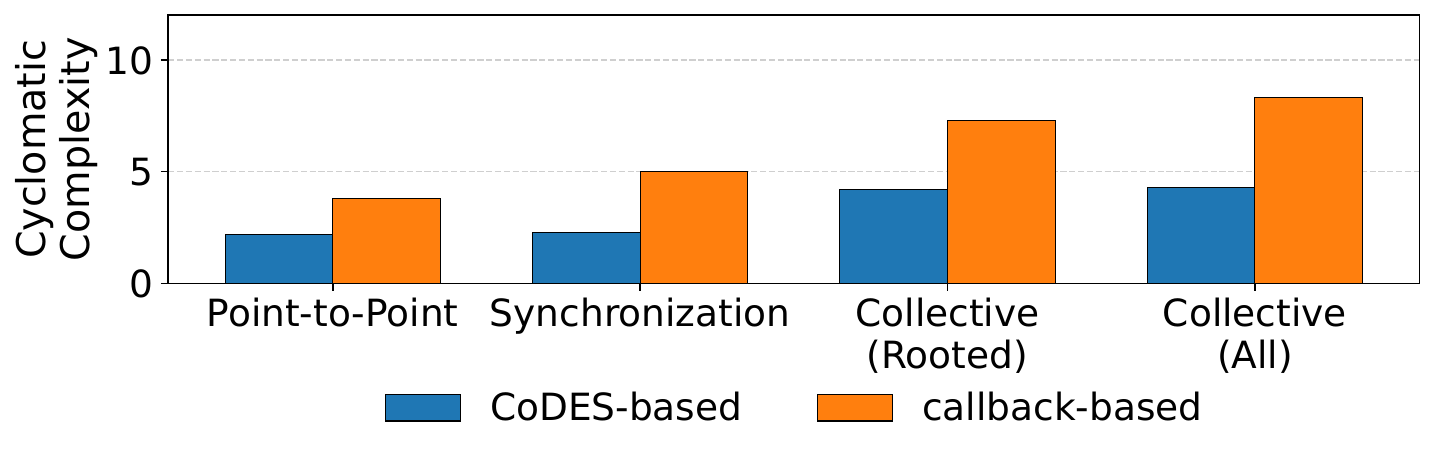}
    \vspace{-0.3cm}
    \caption{Cyclomatic complexity comparison of 20 classical MPI functions in CoDES-based and callback-based implementations. Cyclomatic complexity is a standard software complexity metric that quantifies the number of linearly independent execution paths through a function \cite{mccabe1976cyclomaticcomplexity}.} 
    \vspace{-0.3cm}
    \label{fig:cyclomatric_complexity}
\end{figure}


\subsubsection{Nesting Depth \& Cyclomatic Complexity }
Figure \ref{fig:nesting_depth} compares the nesting depth of 20 representative MPI functions between CoDES-based and callback-based implementations, measured using a commercial code analysis tool \cite{mens2008understanding}. CoDES-based MPI exhibits a nesting depth of 2–6, versus 2–12 for callback-based MPI, yielding a reduction of up to 50\%. Figure \ref{fig:cyclomatric_complexity} shows a similar trend in cyclomatic complexity, where CoDES-based MPI reduces it by 42.1\%–54.0\% compared to the callback-based counterpart, owing to the elimination of redundant branch conditions.

\subsection{AI-Assisted Code Generation Efficiency}
\label{sec:ai_assist_eval}

We evaluate whether CoDES benefits AI-assisted development by prompting DeepSeek-V3 \cite{liu2024deepseek} to implement three typical tasks—Send-Recv, Allreduce, and Alltoallv—under both CoDES-based and callback-based styles with identical descriptions. As shown in Figure \ref{fig:ai-evaluation}, CoDES reduces token consumption by 29\%–87\%, generation time by 60\%–72\%, and error corrections by 27\%–33\%. The wide range in token reduction (29\%–87\%) reflects the task complexity: the simpler Send-Recv task yields a more modest reduction because the callback-based implementation is already relatively straightforward, whereas the structurally more demanding Alltoallv task benefits disproportionately from CoDES.

\subsection{Overhead}
We evaluate overhead along two dimensions: simulation accuracy, measured as the discrepancy between simulated and actual execution traces from large-scale Dragonfly clusters with 1000+ nodes, and runtime cost, measured as execution time and memory usage.

\begin{figure}[htbp] 
\centering

\vspace{-0.3cm}

\subfigure[Simulation time errors relative to the trace baseline.]{
\label{fig:simualtion_error}
\includegraphics[width=0.98\linewidth]{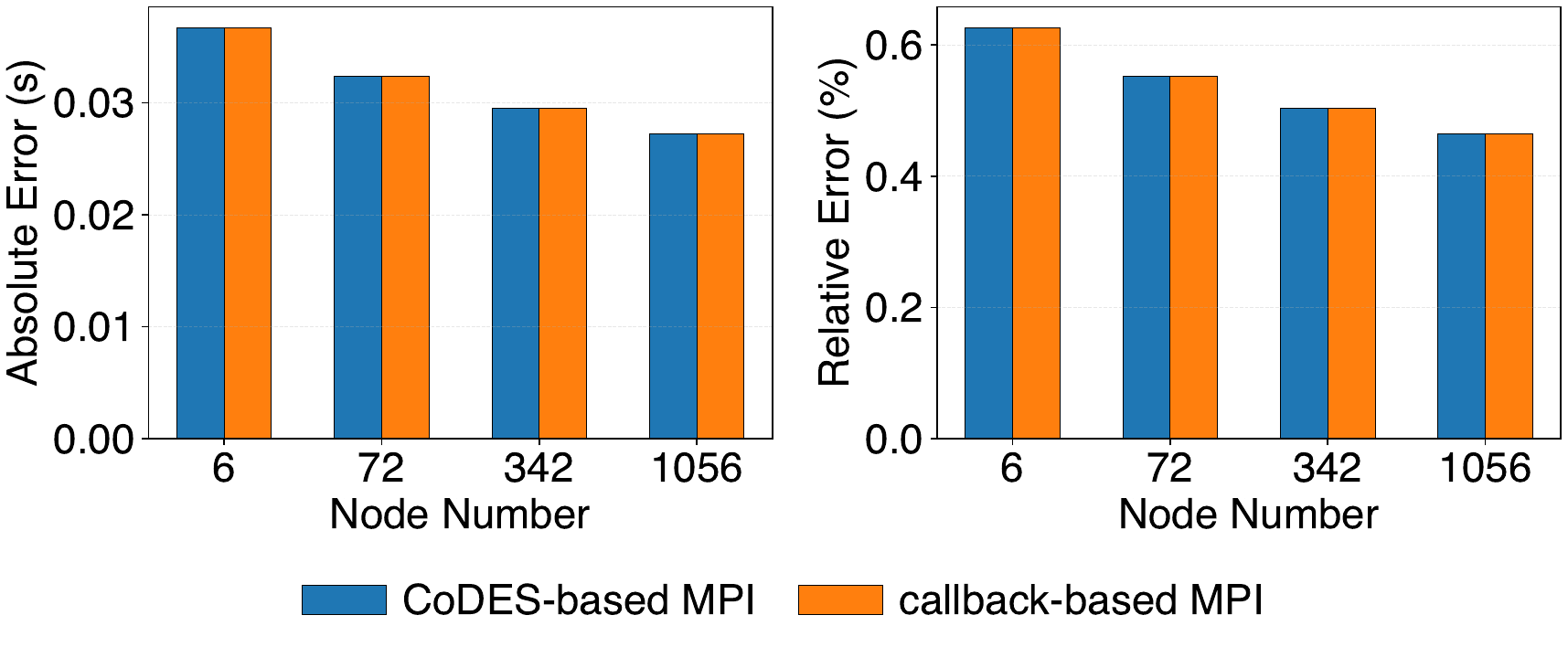}
\vspace{-0.3cm}}

\quad
\vspace{-0.3cm}

\subfigure[Simulation completion time for each MPI function.]{
\label{fig:vtime_progress}

\includegraphics[width=0.98\linewidth]{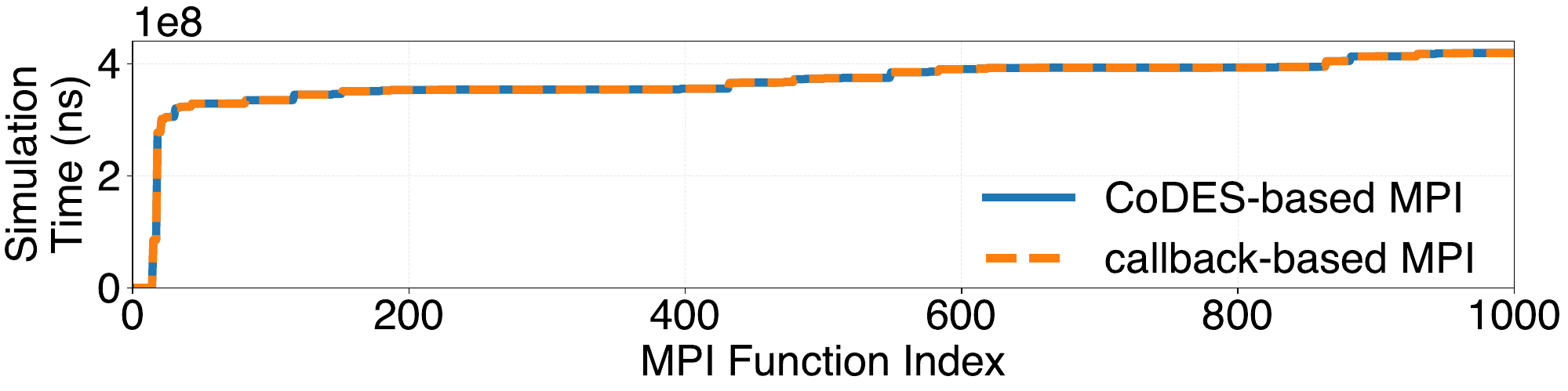}
\vspace{-0.3cm}}

\vspace{-0.2cm}

\caption{Simulation Virtual Time Consistency Between CoDES-based and callback-based MPI.}
\vspace{-0.2cm}
\label{fig:simulation_eval}
\end{figure}


\subsubsection{Simulation Accuracy}
\label{sec:system test}

As shown in Figures \ref{fig:simualtion_error}, the absolute simulation error (defined as the difference between simulated and actual completion time of the application trace) remains within 0.05 seconds and 1\% for both implementations, with highly consistent relative errors. Figure \ref{fig:vtime_progress} further confirms identical per-function execution order and timing, and the CoDES and callback RIP implementations produce byte-identical event traces on the same scenario. Long-run execution under ASan \cite{serebryany2012addresssanitizer} and Valgrind \cite{nethercote2007valgrind} on the 1000+-node configuration reports no memory errors, and stress tests confirm that cancellation, shutdown, and exception propagation behave correctly. Furthermore, under conservative parallel execution, both implementations yield identical simulation time, demonstrating CoDES's correctness under parallel DES.

\begin{table}[t]
\centering
\vspace{-0.35cm}
\caption{Execution time and memory overhead of CoDES-based and callback-based MPI.}
\small
\label{fig:time_rss}
\renewcommand{\arraystretch}{1.15}
\setlength{\tabcolsep}{3.0pt}

\vspace{-0.1cm}

\begin{tabular}{c|c|>{\centering\arraybackslash}m{1.6cm}
>{\centering\arraybackslash}m{1.6cm}|c}
\toprule
\textbf{Metric} &
\textbf{Nodes} &
\centering\textbf{\shortstack{CoDES-\\based MPI}} &
\centering\textbf{\shortstack{callback-\\based MPI}} &
\textbf{Diff.} \\
\midrule

\multirow{4}{*}{\shortstack{Execution Time\\(s)}}
& 6
& \centering 43.83
& \centering 43.73
& \textbf{+0.23\%} \\

& 72
& \centering 58.15
& \centering 58.01
& \textbf{+0.24\%} \\

& 342
& \centering 256.70
& \centering 257.11
& \textbf{-0.16\%} \\

& 1056
& \centering 7216.06
& \centering 7181.21
& \textbf{+0.49\%} \\

\midrule

\multirow{4}{*}{\shortstack{Memory Usage\\(GB)}}
& 6
& \centering 0.0408
& \centering 0.0405
& \textbf{+0.59\%} \\

& 72
& \centering 0.0819
& \centering 0.0813
& \textbf{+0.71\%} \\

& 342
& \centering 0.9181
& \centering 0.9177
& \textbf{+0.05\%} \\

& 1056
& \centering 8.3019
& \centering 8.3017
& \textbf{+0.00\%} \\

\bottomrule
\end{tabular}
\vspace{-0.75cm}
\end{table}




\subsubsection{Execution Time \& Runtime Memory}
As shown in Table \ref{fig:time_rss}, CoDES-based MPI introduces less than 1\% overhead. Similar results are observed in HPCC and RIP. 
In the parallel simulation, execution time and memory usage differ by approximately 0.6\% and 2.6\%, respectively.

\section{Related Work}

\subsection{Programming Paradigms for Event-Driven Systems}

Several approaches have been proposed to restore sequential control flow in
event-driven systems. DepFast \cite{depfast} adopts fibers for cooperative task
management, but fibers require full pre-allocated stacks with high overhead. OMNeT++'s \textit{activity()} model is deprecated for
the same reasons. At the other extreme, Protothreads \cite{protothreads} eliminates
stack overhead but sacrifices local variable preservation across suspension points.
SimPy \cite{zinoviev2024simpy}, and its C++20-coroutine successors simcpp20 \cite{simcpp20} and
libcxxdes \cite{libcxxdes}, target general-purpose DES rather than network simulation, and SimPy's
\textit{yield} cannot propagate through nested calls without spawning a subprocess
per level. General-purpose C++20 coroutine libraries such as cppcoro and folly::coro
schedule resumption on thread-pool executors and preserve neither DES event ordering
nor determinism. Days \cite{li2026days} is a simulator built from scratch in Rust on the actor model. In contrast, CoDES enables incremental adoption across the large existing simulator ecosystem, e.g., ns-3.

CoDES addresses the limitations by applying C++20
stackless coroutines to high-performance network simulation: lightweight like
Protothreads, expressive like fibers, and fully compatible with existing
callback-based components.

\subsection{Network Simulation Systems}

Recent work on network simulation primarily targets execution performance.
Unison \cite{UNISON} and DONS \cite{DONS} accelerate DES through parallel
partitioning and data-oriented scheduling. GeDES \cite{li2026gedes} pioneers
GPU-driven event processing for significant speedup. M3 \cite{m3} leverages
machine learning for flow-level evaluation. These efforts are
\emph{orthogonal} to CoDES: they target execution speed while CoDES targets
development efficiency, and the two objectives are compatible, as CoDES is
experimentally verified to work alongside parallel simulators.

Beyond standalone simulators, SimBricks \cite{li2022simbricks} and SplitSim
\cite{li2024splitsim} compose multiple simulators for modular end-to-end and
large-scale evaluation, while network emulators \cite{ovs, netem, openstack, cao2022turbonet}
provide high-fidelity environments via virtual machines. Both categories rely
on callback-based components. CoDES improves the development experience within
the callback-based components of these frameworks, and its paradigm extends
naturally to any callback-driven simulation scenario.

\section{Discussion}

While CoDES's coroutine abstractions introduce a learning investment, our
evaluation demonstrates that this cost is more than offset by the reduction in
development effort, and that CoDES's sequential structure further benefits
AI-assisted code generation (Section \ref{sec:ai_assist_eval}), demonstrating
that a well-structured programming paradigm remains important even in the era
of AI-assisted development \cite{yu2025genai}.
Although our implementation leverages C++20 coroutines, the CoDES paradigm is
not tied to any specific language or simulator. Any language providing stackless
coroutines with explicit suspension and resumption semantics can serve as the
underlying mechanism. To demonstrate this generality, we implement an Alltoall on OMNeT++ 6.0.3 through a thin CoDES bridging layer, and verify that it produces results identical to an Alltoall built with OMNeT++'s native callback model. 

In parallel simulation, CoDES's correctness relies on the guarantee that no two
concurrent events access the same coroutine frame simultaneously. This property
is ensured by conservative parallel DES \cite{chandy2006distributed, DONS} and
confirmed by our experiments. Extending CoDES
to optimistic approaches \cite{jefferson1985virtual}, which
require coroutine state rollback, remains an open problem. The
CoDES paradigm applies to any callback-based simulator or emulator, as its
design is not specific to ns-3.

\section{Conclusion}
The heavy development workloads in network simulation are often overlooked. We point out that a key cause is the inability of callbacks to naturally simulate blocking network events. We propose CoDES, a coroutine-based programming paradigm that restores sequential workflows for network simulation, and implement it in ns-3 by addressing key system challenges in correctness, functionality, compatibility, and overhead. Results show CoDES reduces code volume and nesting depth by up to 51.4\% and 50\%, respectively.

\section*{Acknowledgment}

We sincerely thank our shepherd and all reviewers for their insightful and constructive feedback. This work is sponsored by the National Key Research and Development Program of China (No. 2024YFE0203900) and National Natural Science Foundation of China (U25B2032, 62572128).

\bibliographystyle{IEEEtran}
\bibliography{reference}

@IEEEtranBSTCTL{BSTcontrol,
  CTLuse_forced_etal       = "yes",
  CTLmax_names_forced_etal = "4",
  CTLnames_show_etal       = "3"
}

@article{felleisen1991expressive,
  title={On the expressive power of programming languages},
  author={Felleisen, Matthias},
  journal={Science of computer programming},
  volume={17},
  number={1-3},
  pages={35--75},
  year={1991},
  publisher={Elsevier}
}

@article{moura2009coroutinetheory,
  title={Revisiting coroutines},
  author={Moura, Ana L{\'u}cia De and Ierusalimschy, Roberto},
  journal={ACM Transactions on Programming Languages and Systems (TOPLAS)},
  volume={31},
  number={2},
  pages={1--31},
  year={2009},
  publisher={ACM New York, NY, USA}
}

@Inbook{ns3-2010,
  title={The ns-3 network simulator},
  author={Riley, George F and Henderson, Thomas R},
  booktitle={Modeling and tools for network simulation},
  pages={15--34},
  year={2010},
  publisher={Springer}
}

@techreport{2015-sst,
  title={Using Discrete Event Simulation for Programming Model Exploration at Extreme-Scale: Macroscale Components for the Structural Simulation Toolkit (SST).},
  author={Wilke, Jeremiah J and Kenny, Joseph P},
  year={2015},
  institution={Sandia National Lab.(SNL-CA), Livermore, CA (United States)}
}

@inproceedings{2016-sstbooksim,
  title={Contention-based congestion management in large-scale networks},
  author={Kim, Gwangsun and Kim, Changhyun and Jeong, Jiyun and Parker, Mike and Kim, John},
  booktitle={2016 49th Annual IEEE/ACM International Symposium on Microarchitecture (MICRO)},
  pages={1--13},
  year={2016},
  organization={IEEE}
}

@article{2008-ns3,
  title={Network simulations with the ns-3 simulator},
  author={Henderson, Thomas R and Lacage, Mathieu and Riley, George F and Dowell, Craig and Kopena, Joseph},
  journal={SIGCOMM demonstration},
  volume={14},
  number={14},
  pages={527},
  year={2008}
}

@inproceedings{DONS,
  title={Dons: Fast and affordable discrete event network simulation with automatic parallelization},
  author={Gao, Kaihui and Chen, Li and Li, Dan and Liu, Vincent and Wang, Xizheng and Zhang, Ran and Lu, Lu},
  booktitle={Proceedings of the ACM SIGCOMM 2023 Conference},
  pages={167--181},
  year={2023}
}

@inproceedings{UNISON,
  title={Unison: A Parallel-Efficient and User-Transparent Network Simulation Kernel},
  author={Bai, Songyuan and Zheng, Hao and Tian, Chen and Wang, Xiaoliang and Liu, Chang and Jin, Xin and Xiao, Fu and Xiang, Qiao and Dou, Wanchun and Chen, Guihai},
  booktitle={Proceedings of the Nineteenth European Conference on Computer Systems},
  pages={115--131},
  year={2024}
}

@inproceedings{opnet,
  title={Network simulations with OPNET},
  author={Chang, Xinjie},
  booktitle={Proceedings of the 31st conference on Winter simulation: Simulation---a bridge to the future-Volume 1},
  pages={307--314},
  year={1999}
}

@incollection{omnet++,
  title={OMNeT++},
  author={Varga, Andras},
  booktitle={Modeling and tools for network simulation},
  pages={35--59},
  year={2010},
  publisher={Springer}
}

@article{glomosim,
  title={Glomosim: A scalable network simulation environment},
  author={Bajaj, Lokesh and Takai, Mineo and Ahuja, Rajat and Tang, Ken and Bagrodia, Rajive and Gerla, Mario},
  journal={UCLA computer science department technical report},
  volume={990027},
  number={1999},
  pages={213},
  year={1999},
  publisher={Citeseer}
}

@inproceedings{ovs,
  title={The design and implementation of open vSwitch},
  author={Pfaff, Ben and Pettit, Justin and Koponen, Teemu and Jackson, Ethan and Zhou, Andy and Rajahalme, Jarno and Gross, Jesse and Wang, Alex and Stringer, Joe and Shelar, Pravin and others},
  booktitle={12th USENIX symposium on networked systems design and implementation (NSDI 15)},
  pages={117--130},
  year={2015}
}

@inproceedings{netem,
  title={Network emulation with NetEm},
  author={Hemminger, Stephen and others},
  booktitle={Linux conf au},
  volume={5},
  pages={2005},
  year={2005}
}

@article{openstack,
  title={OpenStack: toward an open-source solution for cloud computing},
  author={Sefraoui, Omar and Aissaoui, Mohammed and Eleuldj, Mohsine and others},
  journal={International Journal of Computer Applications},
  volume={55},
  number={3},
  pages={38--42},
  year={2012}
}

@article{C++coroutine-1,
  title={C++ 20 coroutines on microcontrollers—what we learned},
  author={Belson, Bruce and Xiang, Wei and Holdsworth, Jason and Philippa, Bronson},
  journal={IEEE Embedded Systems Letters},
  volume={13},
  number={1},
  pages={9--12},
  year={2020},
  publisher={IEEE}
}

@book{C++coroutine-2,
  title={Programming with C++ 20: Concepts, Coroutines, Ranges, and more},
  author={Fertig, Andreas},
  year={2021},
  publisher={Fertig Publications}
}

@inproceedings{depfast,
  title={DepFast: Orchestrating Code of Quorum Systems},
  author={Luo, Xuhao and Shen, Weihai and Mu, Shuai and Xu, Tianyin},
  booktitle={2022 USENIX Annual Technical Conference (USENIX ATC 22)},
  pages={557--574},
  year={2022}
}

@inproceedings {stack-ripping,
  title={Cooperative task management without manual stack management},
  author={Adya, Atul and Howell, Jon and Theimer, Marvin and Bolosky, Bill and Douceur, John},
  booktitle={2002 USENIX Annual Technical Conference (USENIX ATC 02)},
  year={2002}
}

@techreport{edwards2009callbackhell,
  title={Coherent reaction},
  author={Edwards, Jonathan},
  booktitle={Proceedings of the 24th ACM SIGPLAN conference companion on Object oriented programming systems languages and applications},
  pages={925--932},
  year={2009}
}

@inproceedings{m3,
  title={m3: Accurate flow-level performance estimation using machine learning},
  author={Li, Chenning and Nasr-Esfahany, Arash and Zhao, Kevin and Noorbakhsh, Kimia and Goyal, Prateesh and Alizadeh, Mohammad and Anderson, Thomas E},
  booktitle={Proceedings of the ACM SIGCOMM 2024 Conference},
  pages={813--827},
  year={2024}
}

@inproceedings{2019-sigcomm-HPCC,
  title={HPCC: High precision congestion control},
  author={Li, Yuliang and Miao, Rui and Liu, Hongqiang Harry and Zhuang, Yan and Feng, Fei and Tang, Lingbo and Cao, Zheng and Zhang, Ming and Kelly, Frank and Alizadeh, Mohammad and others},
  booktitle={Proceedings of the ACM special interest group on data communication},
  pages={44--58},
  year={2019}
}

@techreport{stark1999rip,
  title={{RFC 2453}: {RIP} Version 2},
  author={Malkin, Gary},
  year={1998},
  month={November},
  institution={Internet Engineering Task Force},
  type={RFC},
  number={2453},
  howpublished={IETF RFC 2453},
  note={\url{https://www.rfc-editor.org/rfc/rfc2453}}
}

@article{dongarra1995MPI,
  title={An introduction to the MPI standard},
  author={Dongarra, Jack J and Otto, Steve W and Snir, Marc and Walker, David and others},
  journal={Communications of the ACM},
  volume={18},
  pages={11},
  year={1995},
  publisher={Citeseer}
}

@article{walker1996mpi,
  title={MPI: a standard message passing interface},
  author={Walker, David W and Dongarra, Jack J},
  journal={Supercomputer},
  volume={12},
  pages={56--68},
  year={1996},
  publisher={ASFRA BV}
}

@article{hughes1989functionalprogramming1,
  title={Why functional programming matters},
  author={Hughes, John},
  journal={The computer journal},
  volume={32},
  number={2},
  pages={98--107},
  year={1989},
  publisher={Oxford University Press}
}

@inproceedings{launchbury1993lazyevaluation,
  title={A natural semantics for lazy evaluation},
  author={Launchbury, John},
  booktitle={Proceedings of the 20th ACM SIGPLAN-SIGACT symposium on Principles of programming languages},
  pages={144--154},
  year={1993}
}

@book{martin2009chaining,
  title={Clean code: a handbook of agile software craftsmanship},
  author={Martin, Robert C},
  year={2009},
  publisher={Pearson Education}
}

@article{DES-intro,
  title={Functional and performance analysis of discrete event network simulation tools},
  author={Musa, Ahmad and Awan, Irfan},
  journal={Simulation Modelling Practice and Theory},
  volume={116},
  pages={102470},
  year={2022},
  publisher={Elsevier}
}

@InProceedings{callback-hell-3,
author="Zamora-G{\'o}mez, Edgar
and Garc{\'i}a-L{\'o}pez, Pedro
and Mond{\'e}jar, Rub{\'e}n",
editor="Hunold, Sascha
and Costan, Alexandru
and Gim{\'e}nez, Domingo
and Iosup, Alexandru
and Ricci, Laura
and G{\'o}mez Requena, Mar{\'i}a Engracia
and Scarano, Vittorio
and Varbanescu, Ana Lucia
and Scott, Stephen L.
and Lankes, Stefan
and Weidendorfer, Josef
and Alexander, Michael",
title="Continuation Complexity: A Callback Hell for Distributed Systems",
booktitle="Euro-Par 2015: Parallel Processing Workshops",
year="2015",
publisher="Springer International Publishing",
address="Cham",
pages="286--298",
isbn="978-3-319-27308-2"
}

@article{josuttis2012c++,
  title={The C++ standard library: a tutorial and reference},
  author={Josuttis, Nicolai M},
  year={2012},
  publisher={Addison-Wesley Professional}
}

@book{klabnik2023cyclic-reference,
  title={The Rust programming language},
  author={Klabnik, Steve and Nichols, Carol},
  year={2023},
  publisher={No Starch Press}
}

@inproceedings {2022-memoryleak,
  title={Resin: a holistic service for dealing with memory leaks in production cloud infrastructure},
  author={Lou, Chang and Chen, Cong and Huang, Peng and Dang, Yingnong and Qin, Si and Yang, Xinsheng and Li, Xukun and Lin, Qingwei and Chintalapati, Murali},
  booktitle={16th USENIX Symposium on Operating Systems Design and Implementation (OSDI 22)},
  pages={109--125},
  year={2022}
}

@techreport{heroux2013HPCG,
  title={A new metric for ranking high-performance computing systems},
  author={Dongarra, Jack and Heroux, Michael A and Luszczek, Piotr},
  journal={National Science Review},
  volume={3},
  number={1},
  pages={30--35},
  year={2016},
  publisher={Oxford University Press}
}

@techreport{karlin2013lulesh,
  title={Lulesh 2.0 updates and changes},
  author={Karlin, Ian and Keasler, Jeff and Neely, J Robert},
  year={2013},
  institution={Lawrence Livermore National Lab.(LLNL), Livermore, CA (United States)}
}

@article{mens2008understanding,
  title={Analysing software repositories to understand software evolution},
  author={Mens, Tom and Demeyer, Serge and D’Ambros, Marco and Gall, Harald and Lanza, Michele and Pinzger, Martin},
  journal={Software evolution},
  pages={37--67},
  year={2008},
  publisher={Springer}
}

@inproceedings{dechev2010ABA,
  title={Understanding and effectively preventing the ABA problem in descriptor-based lock-free designs},
  author={Dechev, Damian and Pirkelbauer, Peter and Stroustrup, Bjarne},
  booktitle={2010 13th IEEE international symposium on object/component/service-oriented real-time distributed computing},
  pages={185--192},
  year={2010},
  organization={IEEE}
}

@article{open-closed-principle,
  title={The open-closed principle},
  author={Martin, Robert C},
  journal={More C++ gems},
  volume={19},
  number={96},
  pages={9},
  year={1996}
}

@techreport{Nishanov2018Fibers,
  author      = {Gor Nishanov},
  title       = {Fibers under the Magnifying Glass},
  institution = {ISO C++ Committee},
  number      = {P1364R0},
  year        = {2018},
  month       = nov,
}

@inproceedings{protothreads,
  title={Protothreads: Simplifying event-driven programming of memory-constrained embedded systems},
  author={Dunkels, Adam and Schmidt, Oliver and Voigt, Thiemo and Ali, Muneeb},
  booktitle={Proceedings of the 4th international conference on Embedded networked sensor systems},
  pages={29--42},
  year={2006}
}

@article{belson2019coroutin-embedded-survey,
  title={A survey of asynchronous programming using coroutines in the Internet of Things and embedded systems},
  author={Belson, Bruce and Holdsworth, Jason and Xiang, Wei and Philippa, Bronson},
  journal={ACM Transactions on Embedded Computing Systems (TECS)},
  volume={18},
  number={3},
  pages={1--21},
  year={2019},
  publisher={ACM New York, NY, USA}
}

@article{mccabe1976cyclomaticcomplexity,
  title={A complexity measure},
  author={McCabe, Thomas J},
  journal={IEEE Transactions on software Engineering},
  number={4},
  pages={308--320},
  year={1976},
  publisher={IEEE}
}

@article{zinoviev2024simpy,
  title={Discrete event simulation: It's easy with simpy!},
  author={Zinoviev, Dmitry},
  journal={arXiv preprint arXiv:2405.01562},
  year={2024}
}

@inproceedings{li2026gedes,
  title={GeDES: GPU-Driven Discrete Event Network Simulator},
  author={Li, Qinyong and Zhao, Zhiwei and Min, Geyong and Wang, Zi and Fu, Luwei},
  booktitle={Proceedings of the 21st European Conference on Computer Systems},
  pages={468--483},
  year={2026}
}

@inproceedings{li2022simbricks,
  title={SimBricks: end-to-end network system evaluation with modular simulation},
  author={Li, Hejing and Li, Jialin and Kaufmann, Antoine},
  booktitle={Proceedings of the ACM SIGCOMM 2022 Conference},
  pages={380--396},
  year={2022}
}

@article{li2024splitsim,
  title={Splitsim: Large-scale simulations for evaluating network systems research},
  author={Li, Hejing and Balasubramanian, Praneeth and Meiers, Marvin and Li, Jialin and Kaufmann, Antoine},
  journal={arXiv preprint arXiv:2402.05312},
  year={2024}
}

@inproceedings{yu2025genai,
  title={Paradigm shift on Coding Productivity Using GenAI},
  author={Yu, Liang},
  booktitle={Proceedings of the 29th International Conference on Evaluation and Assessment in Software Engineering},
  pages={708--713},
  year={2025}
}

@article{chandy2006distributed,
  title={Distributed simulation: A case study in design and verification of distributed programs},
  author={Chandy, K. Mani and Misra, Jayadev},
  journal={IEEE Transactions on software engineering},
  number={5},
  pages={440--452},
  year={2006},
  publisher={IEEE}
}

@article{jefferson1985virtual,
  title={Virtual time},
  author={Jefferson, David R},
  journal={ACM Transactions on Programming Languages and Systems (TOPLAS)},
  volume={7},
  number={3},
  pages={404--425},
  year={1985},
  publisher={ACM New York, NY, USA}
}

@inproceedings{hayamizu2019ceforesim,
  title={CeforeSim: Cefore Compliant NS-3-Based Network Simulator},
  author={Hayamizu, Yusaku and Matsuzono, Kazuhisa and Asaeda, Hitoshi},
  booktitle={2019 IEEE 27th International Conference on Network Protocols (ICNP)},
  pages={1--2},
  year={2019},
  organization={IEEE}
}

@article{cao2022turbonet,
  title={Turbonet: Faithfully emulating networks with programmable switches},
  author={Cao, Jiamin and Liu, Ying and Zhou, Yu and He, Lin and Xu, Mingwei},
  journal={IEEE/ACM Transactions on Networking},
  volume={30},
  number={3},
  pages={1395--1409},
  year={2022},
  publisher={IEEE}
}

@inproceedings{load-balancing-1,
  title={Enabling load balancing for lossless datacenters},
  author={Hu, Jinbin and Zeng, Chaoliang and Wang, Zilong and Zhang, Junxue and Guo, Kun and Xu, Hong and Huang, Jiawei and Chen, Kai},
  booktitle={2023 IEEE 31st International Conference on Network Protocols (ICNP)},
  pages={1--11},
  year={2023},
  organization={IEEE}
}

@inproceedings{anycast-1,
  title={AnyTuner: Optimizing IP Anycast Performance via Strategic BGP Routing Policy},
  author={Zhou, Minyuan and Zheng, Jiaqi and Wang, Congying and Chen, Guihai and Dou, Wanchun},
  booktitle={2025 IEEE 33rd International Conference on Network Protocols (ICNP)},
  pages={1--12},
  year={2025},
  organization={IEEE}
}

@inproceedings{load-balancing-2,
  title={Loom: Switch-based cloud load balancer with compressed states},
  author={Zhang, Jiao and Gao, Yuxuan and Wen, Shubo and Pan, Tian and Huang, Tao},
  booktitle={2021 IEEE 29th International Conference on Network Protocols (ICNP)},
  pages={1--11},
  year={2021},
  organization={IEEE}
}

@inproceedings{anycast-2,
  title={Making multipath TCP friendlier to load balancers and anycast},
  author={Duchene, Fabien and Bonaventure, Olivier},
  booktitle={2017 IEEE 25th International Conference on Network Protocols (ICNP)},
  pages={1--10},
  year={2017},
  organization={IEEE}
}

@article{liu2024deepseek,
  title={Deepseek-v3 technical report},
  author={Liu, Aixin and Feng, Bei and Xue, Bing and Wang, Bingxuan and Wu, Bochao and Lu, Chengda and Zhao, Chenggang and Deng, Chengqi and Zhang, Chenyu and Ruan, Chong and others},
  journal={arXiv preprint arXiv:2412.19437},
  year={2024}
}

@article{reynolds1993discoveries,
  title={The discoveries of continuations},
  author={Reynolds, John C.},
  journal={LISP and Symbolic Computation},
  volume={6},
  number={3--4},
  pages={233--247},
  year={1993},
  publisher={Springer}
}

@book{appel1992compiling,
  title={Compiling with Continuations},
  author={Appel, Andrew W.},
  year={1992},
  publisher={Cambridge University Press}
}

@misc{simcpp20,
  author       = {Felix Sch{\"u}tz},
  title        = {{SimCpp20}: Discrete-Event Simulation in {C++20} Using Coroutines},
  howpublished = {\url{https://github.com/fschuetz04/simcpp20}},
  year         = {2021},
  note         = {Accessed: 2026-08-12}
}

@misc{libcxxdes,
  author       = {Batuhan Nebi},
  title        = {{libcxxdes}: A {C++20} Coroutines-Based Discrete-Event Simulation Library},
  howpublished = {\url{https://github.com/jnbrq/libcxxdes}},
  year         = {2023},
  note         = {Accessed: 2026-08-12}
}

@inproceedings{li2026days,
  title={Days: Discrete-Event Network Simulation on Steroids},
  author={Li, Baochun},
  booktitle={IEEE INFOCOM 2026-IEEE Conference on Computer Communications},
  pages={1--10},
  year={2026},
  organization={IEEE}
}

@inproceedings{serebryany2012addresssanitizer,
  title={$\{$AddressSanitizer$\}$: A fast address sanity checker},
  author={Serebryany, Konstantin and Bruening, Derek and Potapenko, Alexander and Vyukov, Dmitriy},
  booktitle={2012 USENIX annual technical conference (USENIX ATC 12)},
  pages={309--318},
  year={2012}
}

@article{nethercote2007valgrind,
  title={Valgrind: a framework for heavyweight dynamic binary instrumentation},
  author={Nethercote, Nicholas and Seward, Julian},
  journal={ACM Sigplan notices},
  volume={42},
  number={6},
  pages={89--100},
  year={2007},
  publisher={ACM New York, NY, USA}
}

\end{document}